\documentclass[12pt,a4paper]{article}
\usepackage[T1]{fontenc}
\usepackage{indentfirst}
\usepackage[english]{babel}
\usepackage{amsfonts,latexsym}
\usepackage{amsthm}
\usepackage{euscript}
\usepackage{amsmath}
\usepackage{color}
\usepackage{graphicx}
\usepackage{caption}
\usepackage{subcaption}
\numberwithin{equation}{section}
\usepackage{ucs}
\usepackage[utf8x]{inputenc}
\usepackage{amsfonts}
\usepackage{graphicx}
\usepackage{fancybox}
\usepackage{tikz}
\usepackage{rotating}

\newfont{\bcb}{msbm10 scaled 1200}
\newfont{\bcc}{msbm10}

\title{Aggregated moving functional median in robust prediction of hierarchical functional time series  - an application to forecasting web portal users behaviors}
\author{Daniel Kosiorowski$^1$, Dominik Mielczarek$^2$,
\\ Jerzy P. Rydlewski$^2$(corresponding author)}
\begin{document}
\maketitle
\begin{center} 
$^1$\textit{Cracow University of Economics, Department of Statistics, Poland}
\\ $^2$\textit{AGH University of Science and Technology, Faculty of Applied Mathematics, Krakow, Poland;} 

\end{center}





\begin{abstract}
In this article, a new nonparametric and robust method of forecasting hierarchical functional time series is presented. The method is compared with Hyndman and Shang's method with respect to their unbiasedness, effectiveness, robustness, and computational complexity. Taking into account results of the analytical, simulation and empirical studies, we come to the conclusion that our proposal is superior over the proposal of Hyndman and Shang with respect to some statistical criteria and especially with respect to robustness and computational complexity. An empirical usefulness of our method is presented on example of  management of a certain web portal divided into four subservices. An extensive simulation study involving hierarchical systems consisted of FAR(1) processes and Wiener processes has been conducted as well.  
\end{abstract}
\textbf{keywords:}
\\ Functional Data Analysis, Hierarchical Time Series Forecasting, Moving Functional Median, Robust Forecasting, Reconciliation of Forecasts, Bottom-Up Method
\section{Introduction}
\label{intro}
In the area of econometric modeling we often meet a system of dynamic economic phenomena with imposed structure of hierarchy of its components. As a macroeconomic example, let us consider an export of a country divided into product and service groups and subgroups and additionally into its geographical regions or its administration units. As an microeconomic example, let us take an on-line shop selling cosmetics divided into the so-called target groups and additionally divided into geographical regions, sex, or age groups.\\
A reconciliation of forecasts obtained on different levels or knots of hierarchy has been considered in the statistical as well as econometric literature (see \cite{Hyndman2011,Kahn,Kohn,Weale}) and concerns combining forecasts obtained at different levels or knots of the hierarchy. Particularly, by the reconciliation we mean the desired property of a predictor that sum of forecasts obtained in knots at some level equals to the forecast obtained in the relevant knots on the upper level.\\ The difference between forecasts obtained at lower levels and the upper levels might be a result of different measurement methods, different methodologies, or measurement precision applied for different levels of the hierarchical structure.\\
It is noteworthy to note that in economics there are many phenomena which may be described as functions of certain continuous variables, for example, utility or yield curves, capital flows, the Internet traffic \cite{Horvath,Bocian,StatPap,Ramsay}, or air pollution intensity within day and night. 
Regarding the last issue, as the air pollution has a negative impact on human health, the precise air pollution forecast may be used to maximize a summarized utility of some local community over a
certain period \cite{cejeme} and thus to obtain a maximization of social welfare. It is a result of the fact that utility function over a certain period is a function of, i.a., costs related to quality of forecasting. 
In the context of the Internet service management, one may desire a certain kind of robustness of a maximization of utility related to security. We mean here the robustness of a security system to "not rapid but systematic" intrusions into the system.  \\ An evolution in time of such phenomenon may naturally be described by means of functional time series, that is, family of functions indexed by time. If additionally the time series exhibit a hierarchy (i.e., e.g. a web portal divided into sub-services), it is natural to consider a hierarchical functional time series (HFTS) \cite{Shang}. The HFTS is a functional time series grouped with respect to relevant levels of hierarchy. Forecasting a HFTS means to prepare forecasts jointly for all knots at all levels. \\
A robust HFTS forecasting method generally mean a method, which produce a "good" prognoses \emph{for the whole hierarchical structure} despite a small perturbation of the underlying model or a small contamination of an underlying dataset. The contamination comprises of a replacement of a certain fraction of the considered dataset by shape or magnitude functional outliers (see \cite{Riani}). In a context of a merit interpretations of HFTS, it is worth discriminating between structural changes affecting the whole structure of the hierarchical system (as effects of financial crashes, extraordinary political events) and "usual" outliers, which also may be treated as "packages" of important information but of a smaller scale/merit importance.
Note that after obtaining a prognosis on each level of the hierarchy, usually a natural problem arises--how to properly adjust information obtained on each level in order to obtain a globally re-conciliated prognosis for all levels?\\
In this context, several HFTS forecasting methods have been proposed in a literature.
\\ Bottom-up method involves making base forecasts (in knots at the lowest level of the hierarchy) at the beginning. Then, an aggregation of the obtained forecasts takes place in order to obtain forecasts for knots at upper levels of the hierarchy. A specific form of the aggregation relates to historical data on ratios of sizes of the considered knots or levels. Top-down method draws forecasting the series in knots at the top level of the hierarchy. The disaggregation is subsequently made. The aforementioned methods are often combined, that is, the forecast for knots at some intermediate level of hierarchy is made and then the aggregation produces forecasts on higher levels of the hierarchy, whereas disaggregation gives forecasts on lower levels.\\
It should be stressed that if we make forecasts at all levels of hierarchy independently, then a sum of the forecasts on the lower level very often do not sum up to the forecast on the upper level \cite{Hyndman2011}. Note that all the aforementioned methods do not take into account a dependency structure (e.g. a correlation structure) between the knots at levels of the hierarchy.
\\ In the study by Shang and Hyndman \cite{Shang}, authors have considered re-conciliated forecasts for knots at all the levels of the hierarchy, extending a method proposed by Hyndman et al.  \cite{Hyndman2011} to a functional time series case. Their approach takes into account a known dependency structure between levels and uses forecasts dispersions estimates calculated from empirical data. The conducted reconciliation is based on an application of a certain form of the generalized least squares method, which is well known to be a non-robust method.
They evaluated uncertainty of forecasts using a relevant bootstrap method, which is appropriate for time-dependent observations. \\
After conducting several simulations and empirical studies, we aimed to propose a conceptually simple alternative approach, which is robust to functional outliers and which is not as computationally intensive as theirs.\\
We use an aggregated moving functional median--a functional median of moving functional median in order to obtain a re--conciliated prediction for all considered levels of hierarchy. Our forecast method seems to be robust to shape as well as magnitude outliers and is computationally tractable.
\\ The primary objective of this paper is to propose an effective, robust, and computationally tractable method of forecasting of a HFTS. We also present a successful application of the proposal to the real data, namely monitoring the Internet service divided into four subservices. The remainder of the paper is organized as follows: Section 2 presents some theoretical notions and results we use in our considerations. Section 3 presents our proposal. Section 4 shows some properties of our proposal. Section 5 presents a simulation study, where our method has been applied to an artificial hierarchical systems consisted of functional autoregression processes of order one (FAR(1)) and Wiener processes. Section 6 presents an application of our method to analysis of the web portal users' behaviors, as we would like to test our forecast method on real observed data.
Finally, we end this paper with some conclusions presented in Section 7. An implementation of the method together with the considered datasets is available in free R package \textit{DepthProc} \cite{Kos1}.
\section{Theoretical framework}
\label{sec:2}
Following description of the problem presented in the preceding section, it is advantageous to consider some economic phenomena as functions. The functional data analysis (FDA) methodology has been developed recently (for details see \cite{Bosq,Krzysko,Horvath,Ramsay,Shang}). 
Following the previously mentioned studies, we do consider random curves as real valued functions on $[0, T]$, where $T$ is known. 
The random curves are elements of the space $L^2([0,T])$, a separable Hilbert space with the inner product $\left<x,y\right>=\int x(t)y(t)dt$, and equipped with the Borel $\sigma-$algebra \cite{Bosq}.
Functional time series is a time-indexed series of functional random variables. In empirical studies, we often observe trajectories of functional series at a different "levels", e.g., electricity demand within day and night for a single household, town, and district or an air pollution in a town, a single state, and for a group of states. Such data impose a natural hierarchical structure on a studied phenomenon.
\subsection{Functional depth}
The functional depth concept \cite{Nagy,Nieto} is crucial in our considerations, as we use the concept to define the relevant functional median.
\\ To define modified band depth (MBD, see \cite{LopezRomo}) of curve $x$ with respect to functional sample $X^N$, we need to introduce a measure of the set in the domain that measures centrality of the function with respect to the considered sample. 
We consider a sample of $N$ functions $X^N=\{x_1,...,x_N\}$. Let us define $A(x;x_{i_1},x_{i_2})=\{t\in[0,T]:  \min_{r=i_1,i_2} x_{r}\leq x(t)\leq \max_{r=i_1,i_2} x_{r} \}.$ 
Note that for computational reasons, we consider only a case of bands limited by two functions $x_{i_1}$ and $x_{i_2}$; however L\'opez-Pintado and Romo \cite{LopezRomo} define the MBD for greater number of limiting curves. MBD measures a proportion of "time", when $x$ is in the corresponding band, namely
\begin{equation}
MBD(x|X^N)=\frac{2}{N(N-1)}\sum_{1\leq i_1< i_2\leq N} \frac{\lambda(A(x;x_{i_1},x_{i_2}))}{\lambda([0,T])},
\end{equation}
where $\lambda(\cdot)$ denotes a Lebesgue measure.
The deepest curve in the MBD sense may be different in shape from the majority of the curves in the sample, but is close in the space $L^2$ norm to the mean.
\\ Generalized band depth (GBD, see \cite{Lopez}) of curve $x$ with respect to functional sample $X^N$ estimates the curves' frequency of being in the center in the following sense. 
\begin{equation}
GBD(x|X^N)=\frac{2}{N(N-1)}\sum_{1\leq i_1< i_2\leq N} \frac{CI(x;x_{i_1},x_{i_2})}{\lambda([0,T])},
\end{equation}
where $CI(x;x_{i_1},x_{i_2})=\max_{t_S}\{\lambda(t_S) : \min_{r=i_1,i_2} x_{r}\leq x(t)\leq \max_{r=i_1,i_2} x_{r}, \forall t\in t_S \},$ where $t_S\subset[0,T]$ is a compact interval. Thus, $CI(x;x_{i_1},x_{i_2})$ is the longest consecutive interval, for which $x$ belongs to the band limited by functions $x_{i_1}$ and $x_{i_2}$.
\\ Note that the following inequality is always true: $GBD(x|X^N)\leq MBD(x|X^N)$. As can be deduced from the preceding formulas, $GBD$ enables to distinguish functions differing in shape, while $MBD$ enables to distinguish functions differing in magnitude. It is also worth to notice, that functions which are close to the centre of the sample and not differing in shape have rather large both $GBD$ and $MBD$ values, while functions which are close to the centre of the sample but differing in shape have rather small $GBD$ value in comparison to $MBD$ value. Large $GBD$ value is connected with functions of common shape. The more irregular in shape are the curves in the sample, the less $MBD$ is appropriate, and the more $GBD$ is appropriate. We recommend to undertake a pre-analysis of kind of the analyzed functions in order to decide, which functional depth is the most relevant. Comparison of various functional depths obtained for a given curve w.r.t functional sample may give a substantial information, as well.
\\ Afterwards, the nested regions for depth can be constructed, that is, consider i.e. $GBD(x|X^N)\geq \alpha$. The functional median with respect to the considered functional depth is the most central observation. Assume now, we have $N$ functions at our disposal, that is,  $X^N=\{x_i(t), i=1,2,...,N \}$ and $t\in [0,T].$
Let $FD(y|X^N)$ denote sample functional depth of function $y(t)$ with respect to sample $X^N$ (in the paper we consider only $FD=MBD$ or $FD=GBD$). A sample median w.r.t. the considered functional depth can be defined as 
$$MED_{FD}(y|X^N)={\mathop{\arg \max }}_{i=1,...,N} FD(x_i|X^N).$$
If more than one function achieves the depth maximum value, the functional median is defined as mean of functions placed in a convex hull of such functions \cite{Kos1}.
Note that as different functional depths can be considered, the different functional medians could be obtained.
Subsequently, we use the moving functional median  
$$\hat{x}_{n+1}(t)=MED_{FD}(W_{n,k})$$ 
where $W_{n,k}$ is a moving window of length $k$ with an end in a moment $n$, that is, $W_{n,k}=\{x_{n-k+1},...,x_n\}.$
\subsection{A reference approach to HFTS forecasting}
Although there are several ad hoc and more systematic approaches to the HFTS forecasting known from the literature, we have taken Shang and Hyndman's method as a reference approach mainly because of its mathematical elegance, good statistical properties and comprehensiveness. Their approach descends from their previous studies (see, i.e. \cite{Shang}) which take into account the complete hierarchy.
The method has been recently modified using a generalized exponential
smoothing technique for the most disaggregated functional time series in order to obtain a more robust predictor (for details see \cite{SIT}).
Shang and Hyndman's method of mortality rates forecasting uses a functional principal component analysis in order to perform functional principal component regression at each level, where the times series forecast of the principal component scores are obtained with an effective but nonrobust univariate time series forecasting method, namely Hyndman and Shang \cite{HS2009} method (see also \cite{KosFuncReg}). 
The re-conciliated forecast is a certain generalized least squares estimator of the forecasts obtained for the functional time series at each level of the considered hierarchy. 
The structure of the hierarchical time series at each moment is modeled then with the matrix equation, which takes into account all time series at all levels of hierarchy, notably a vector of the most disaggregated series coming from nodes (knots) at the lowest level of the hierarchy. 
As a result the approach is endowed with a direct internal mechanism of reconciliation of forecasts.
For a more detailed description of Shang and Hyndman method see their paper \cite{Shang}.
Among the properties of their proposal there is an aggregate consistency, namely the forecasts satisfy an aggregation constraints and are mean-unbiased (see \cite{Shang}). Nevertheless, the method we find rather computationally demanding and relatively sophisticated (sparse linear least squares issue, generalized inverses of large matrices). Our method is deprived of the latter disadvantages.
\section{Our proposal} 
\label{sec:3}
In the first step, we calculate the moving functional median related to the GBD, MBD, or another functional depth (FD) for each knot at the lowest level of hierarchy, that is, $$MED_{FD}(W_{n,k})$$ is calculated.
For instance, in our empirical example for the Internet service (web portal) we compute for each of four sub-services (hereafter denoted $SS_1$, $SS_2$, $SS_3$, and $SS_4$) at moment $n$ a functional median from a moving window of length $k$ with respect to the functional depth $GBD$ and substitute: $$\hat{x}^{SS_i}_{n+1}(t)=MED_{GBD}\{x^{SS_i}_{n}(t),x^{SS_i}_{n-1}(t),...,x^{SS_i}_{n-k+1}(t)\},$$ where $i=1,2,3,4$.
\\In the second step, we calculate for the lowest but one level of the hierarchy, a sum of relevant functional medians calculated at corresponding moments for knots in the first step.
\\ We repeat the second step moving up the hierarchy until we calculate the sum of functional medians for the top level of the hierarchy.
\\ In our empirical example, the second step is the last one, and finally we obtain a forecast for $n=k,...,365$
\begin{equation}
\hat{x}_{n+1}(t)=\sum_{i=1}^4 \hat{x}^{SS_i}_{n+1}(t).
\label{suma_med}
\end{equation}
Note that the resulting sum of functional medians does not necessarily equal a functional median of sums of the corresponding functions.
\\ The equality does not hold even for sequences of natural numbers configured into a hierarchy. Let us take two knots $x_1=(0,1,1)$ and $x_2=(1,0,1)$ at the bottom level and it's sum $x_1+x_2=(1,1,2)$ on the second and equivalently top level of the hierarchy. The following statement is true\\
$MED\{(1,1,2)\}=1 \neq 2= MED\{(0,1,1)\}+MED\{(1,0,1)\}.$
Generally, we obtain
\begin{equation}
MED\{(x_1,...,x_m)+(y_1,...,y_m)\}\neq MED\{(x_1,...,x_m)\} +MED\{(y_1,...,y_m)\},
\label{med}
\end{equation}
where $x_i$ and $y_i$ are numbers.
Property (\ref{med}) is not a case of functional mean for equinumerous sets, where equality obviously holds, provided functional mean exists. 
\\ Therefore, the hierarchical structure of the data is accounted in the process of computing forecasts for the upper levels and consequently in the process of aggregation of the functional medians upward the considered hierarchy. Translation of a single functional observation at the same hierarchy level from one knot to another can easily alter the outcome (see inequality (\ref{med})). When the hierarchical structure is quite plenteous, the fact is even more evident.
\\ Obviously, the resulting sum of functional medians (\ref{suma_med}) does not necessarily equal a sum of functional medians of all functions combined either.
\subsection{Uncertainty evaluation}
In general terms, as far as the authors know, an uncertainty evaluation in the HFTS setup is still an open problem. To the best of our knowledge, there are no satisfying theoretical solutions to the problem of confidence bands determining.
In their paper Aue et al. (2015) \cite{aue} (see Section 5.2 of their article) proposed an algorithm for determining uniform prediction bands.
They calculated the residuals employing sample functional principal components (FPC) from the entire sample. The confidence interval for residuals was calculated then. Assuming that the residuals are approximately stationary, the prediction band was determined. The forecasting method that produced smaller prediction band was considered to be better.
\\ Shang and Hyndman (2017) \cite{Shang} adapted the method of Aue et al. (2015) \cite{aue} for computing uniform and pointwise prediction intervals. 
However, Aue et al. (2015) calculated the standard deviation of the residuals and then constructed the final confidence band, while Shang and Hyndman (2017) sampled with replacement to obtain a series of bootstrapped forecast errors and, after that, constructed the final confidence band.
\\ Similarly, Shang (2018) \cite{shang2018} stated that resampling methodology, especially
bootstrapping, turned out to be the only practical method
in estimating the variability associated with functional estimator and constructing proper confidence intervals.
\\ Principal component scores (PCS) are commonly considered as substitutes of original functional time series, therefore PCS imitate the dependence structure of the original FTS.
For this reason, a functional time series is transformed into a family of one-dimensional PCS series, and then a maximum entropy bootstrap methodology of Vinod and de Lacalle \cite{Vinod}, implemented in their \textit{meboot} R package is used \cite{shang2018}. 
\\ This is an appealing simplification of the problem, but it certainly deprives a researcher of the large amount of information on the analyzed functional time series. Following \cite{Lopez_scale} (see also \cite{Tarabelloni}), we propose, therefore, using functional boxplots and adjusted functional boxplots, where sizes of boxes and $\alpha-$central regions are considered. The idea of relevant bootstrap for functional time series for the forecast uncertainty approximate evaluation is fulfilled in this way.   
\\ Similarly to Aue et al. (2015) approach, we reckon that  the forecasting method that produces narrower $\alpha-$central regions is better.
\\ Alternatively, a comparison of functional time series predictor "effectiveness" may be carried out throughout the comparison of speeds of expansion of these central regions as a functions of $\alpha$ (scale curve, see \cite{Lopez_scale}).
\section{Properties of our proposal} 
\label{sec:4}
\subsection{Unbiasedness of a forecast in a single knot}
Let us consider i.i.d. functional observations in a knot at the lowest level of the hierarchy. 
Note that it is not advisable to consider point-wise properties of the predictors. Moreover, they always depend on the data-driven intervals $A(x;x_{i_1},x_{i_2})$ or $CI(x;x_{i_1},x_{i_2})$ in equations (1) or (2) that are taken into account when calculating the $MBD$ or $GBD$ depth, respectively.
\\ We usually do not know the true distribution on the $L^2[0,T]$ space, from which our data come from, so we even cannot straightly assume that the functional mean does exist. As a result of which considering a functional counterpart of an usual mean-biasedness we find inappropriate. 
However, if the estimator is mean-unbiased for knots at the bottom level of the hierarchy, then it is mean-unbiased for knots at the higher levels of the hierarchy. Point-wise unbiasedness, which may ignore important properties of curves, seems to be even worse solution. Instead of this, we focus our attention on the median-unbiasedness. The idea for one-dimensional setting has been reactivated in Brown's paper \cite{brown}: an estimate of a parameter is median-unbiased if, for fixed parameter value, the median of the distribution of the estimate is at the parameter value. In other words, the estimate overestimates just as often as it underestimates. Optimal median-unbiased estimators have been considered by Pfanzagl \cite{pfan79} and asymptotic efficiency of median-unbiased estimates have been examined by Pfanzagl \cite{pfan70}. 
In the considered functional setup, the functional depth is selected and hereunder the median induced by this depth can be computed. Note that the functional median calculated with respect to the selected functional depth is intrinsically a median-unbiased estimator (median with respect to the same functional depth). The functional medians induced by commonly used depth exist for a wide class of processes, in contrary to the existence of functional mean.  
\subsection{Consistency}
Mosler and Polyakova \cite{MoslerPolyakova} note that their $\Phi$-depths (considered generalized band depth clearly does not belong to the class) cannot be meaningfully related to a data generating probability model, and no consistency or other asymptotic results are straightly available.
Fortunately, consistency of the band depth-based median estimator would result from the Gijbels and Nagy's article \cite{gijbels_nagy}, where the authors solved several consistency problems for nonintegrated depth-class estimators (our estimator does not belong to the class considered in their paper) with adjusted band depth for which it is possible to guarantee a uniform consistency result.
\\ However, we use a modified band depth ($MBD$) depth of L\'opez-Pintado and Romo \cite{LopezRomo} and
generalized band depth ($GBD_I$) of L\'opez-Pintado and J\"ornsten \cite{Lopez} in order to define an aggregated median estimator. $MBD$'s consistency, when a moving window length goes to infinity, has been recently proved in Nagy et al. \cite{Nagy}, under some reasonable assumptions. 
\\Roughly speaking, a univariate depth of function $x(t)$ w.r.t the corresponding marginal distribution of the given probability can be calculated. Afterwards, we can compute average of the univariate depths ($D$) as the integrated depth of $x$. Precisely, following Nagy et al. \cite{Nagy} (see their Definition 2.3, p. 100) for $P\in \mathcal{P}(\mathcal{C}([0,1]))$, which is a probability measure on a measurable space $\left( \Omega, \mathcal{F}\right)$, and for a continuous function $x\in \mathcal{C}([0,1])$, the integrated depth of function $x$ w.r.t. $P$, is defined by the formula
$$FD(x;P,D)=\int_0^1D(x(t),P_t)d\mu(t).$$
In order to show that $GBD$ does belong to the class of integrated functional depths of Nagy et al. \cite{Nagy}, following the authors notation, let us consider the univariate depth for $J=2$, as we consider in (2) only the bands limited by two functions. Let also $v\in\mathbf{R}$ and $Q\in\mathcal{P}(\mathbf{R})$ be a collection of all probability measures on real numbers $\mathbf{R}$, and let independent random variables $V_1,V_2 \sim Q$. We define a univariate depth with the formula
$$D_{GBD}(v;Q)=P\left(v\in [\min \{V_1,V_2\}, \max \{V_1,V_2\}]\cdot \mathbf{1}_{CI(v;V_1,V_2)} \right).$$
In order to prove, that $GBD$ defined with (2) is indeed in the class of integrated functional depths, we need to verify the properties $(D_1)-(D_7)$ of Nagy et al. \cite{Nagy} for the for real-valued ($K = 1$ is substituted) functional depth $D_{GBD}$, assuming that measure $Q$ is absolutely continuous. However, we replace their weaker condition ($D_3^W$) instead of their condition ($D_3$) on the condition list below. The properties are:
\\ ($D_1$) Affine invariance: For any nonzero $A\in \mathbf{R},$ and for any $b,v\in \mathbf{R}$ and $V$:
$$D(v;Q_V)=D(Av+b;Q_{AV+b}).$$
\\($D_2$) Maximality at center: If the distribution of $V$ is halfspace symmetric around $v^* \in \mathbf{R}$, then $D(v;Q_V)$ attains its maximum at $v^*$. 
\\ ($D_3$) Decreasing along rays: If a maximum depth $D$ is attained at $v^*\in\mathbf{R}$, then for every  $v\in\mathbf{R}$ and $\gamma\in[0,1]$
the following inequality holds
$$D(\gamma v^*+(1-\gamma)v;Q)\geq D(v;Q).$$
($D_4$) Vanishing at infinity: $$\lim_{|v|\to \infty}D(v;Q)=0,$$ where $|v|$ denotes the Euclidean norm on $\mathbf{R}$.
\\ ($D_5$) Upper semicontinuity of $D$ as a function of $v$:
For all $v\in \mathbf{R}$ and $\lim_{\delta\to \infty}v_{\delta}\to v$ the following formula holds
$$\limsup_{\delta\to \infty}D(v_{\delta};Q)\leq D(v;Q).$$
\\ ($D_6$) Weak continuity of $D$ as a functional of $Q$: For all $Q_{\delta}\buildrel w\over\longrightarrow Q, \textrm{ if }\delta\rightarrow \infty$ the following formula holds
$$\sup_{||v||\in\mathbf{R}}|D(v;Q_{\delta})-D(v;Q)|\longrightarrow 0,\textrm{if } \delta\to\infty.$$
\\ ($D_7$) Measurability:
The mapping
$$D:\mathbf{R}\times \mathcal{P}(\mathbf{R})\to [0,1]:(v;Q)\mapsto D(v;Q)$$
is jointly Borel measurable and $D(\cdot;Q)\not\equiv 0$ for all $Q\in \mathcal{P}(\mathbf{R})$.
\vspace{2mm}
\\ \textbf{Remark 1:} The property $(D_1)$, namely affine invariance, is trivial.
\vspace{2mm}
\\ \textbf{Lemma:} If $Q$ is absolutely continuous,then $D_2$ is satisfied for the univariate $D_{GBD}$ depth.
\begin{proof} We are following the lines of Nagy et al. \cite{Nagy} Theorem A.11 adjusted to the considered univariate $D_{GBD}$ depth. Let $F$ be the continuous distribution function of measure $Q$. For any $v\in \mathbf{R}$ we have
$$ D_{GBD}(v;Q)=P\left( v\in [\min \{V_1,V_2\}, \max \{V_1,V_2\}]\cdot \mathbf{1}_{CI(v;V_1,V_2)} \right)=$$
$$=1-P\left( v\notin [\min \{V_1,V_2\}, \max \{V_1,V_2\}]\cdot \mathbf{1}_{CI(v;V_1,V_2)} \right)=$$ 
$$=1-P\left( v\leq \min \{V_1,V_2\}\cdot \mathbf{1}_{CI(v;V_1,V_2)} \vee v\geq \max \{V_1,V_2\}]\cdot \mathbf{1}_{CI(v;V_1,V_2)} \right)=$$
$$=1-\left(F(v)^2+(1-F(v))^2 \right).$$
Hence $D_{GBD}(v;Q)$ attains its maximal value for $F(v)=\frac12$, namely at the median of $Q$. As this is a one-dimensional distribution, it is halfspace symmetric only around the median (see Nagy et al. \cite{Nagy} p. 98).
\end{proof}
\vspace{2mm}
\textbf{Remark 2:} 
If measure $Q$ is absolutely continuous, then following the lines of proof of \cite{LopezRomo} Theorem 1, we obtain that $(D_3)$ is satisfied. Properties $(D_4)$ and $(D_5)$ are satisfied as well (see L\'opez-Pintado and Romo \cite{LopezRomo}, Theorem 1). 
\vspace{2mm}
\\ \textbf{Remark 3:} As our $D_{GBD}$ is a restricted version of Nagy et al.'s $D_B^2$ depth (see \cite{Nagy} for details), then assuming that $Q$ is absolutely continuous and following the lines of their Theorems A.10, and A.12, we get the properties $D_7$ and $D_6$, respectively. We now take advantage of their Remark A.14 to state that our $GBD$-based estimator is a strong universal consistent estimator (($sC_2$) in Nagy et al. \cite{Nagy}) in knots at the bottom level of the hierarchy.
\vspace{2mm}
\\ \textbf{Remark 4:} Consistency properties in any knot for each level of the hierarchy are obtained for a MBD-based median estimators under the same reasonable conditions as well. This time, it is a direct result proved in Nagy et al. article \cite{Nagy} (see their subsection A.3, p. 121 and Remark A.14, p. 123).
\vspace{2mm}
\\ \textbf{Remark 5:} As in the considered hierarchical structure,  the random variable of each non-bottom knot is a sum of random variables of knots below, then the functional median based estimator is consistent for knots at any level of the hierarchy.
\vspace{2mm}
\\ Our further interest in statistical inference goes to quantities derived from the functional sample. The interest goes to the maximizer of the functional sample depth, and a question is whether for large sample sizes $n$, this maximizer is close to the set of maximizers of the population version of the functional depth function. We use Nagy et al. study \cite{Nagy} (see Section 4.2) to obtain an existence, measurability, and continuity of the applied functional median with respect to the $GBD$ and $MBD$. 
The sample versions of the $GBD$ and $MBD$ depths are strongly consistent as well (see Remark A.14, p. 123).

\subsection{Robustness and "effectiveness"}
\begin{figure*}
\includegraphics[width=\textwidth]{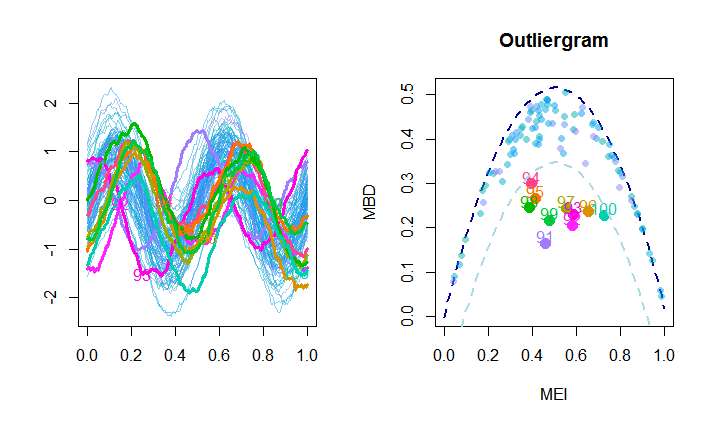}
\caption{Data contaminated with 10\% shape outlying observations on the left and an outliergram for data contaminated with 10\% shape outlying observations on the right.}
\label{fig:2}
\end{figure*}

\begin{figure*}
\includegraphics[width=\textwidth]{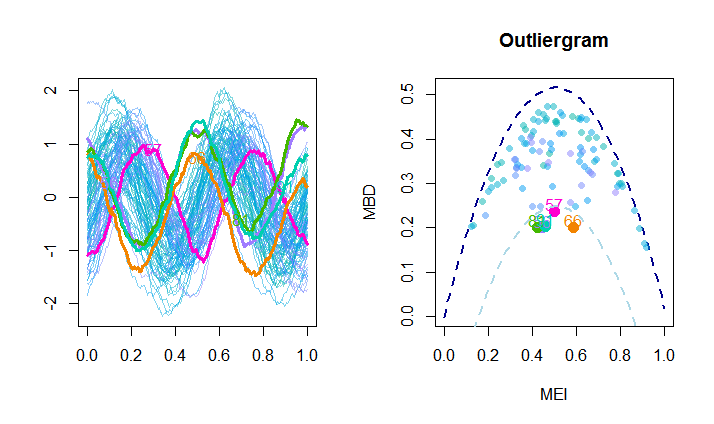}
\caption{Data contaminated with 45\% shape outlying observations on the left and an outliergram for data contaminated with 45\% shape outlying observations on the right.}
\label{fig:3}
\end{figure*}

Although Shang and Hyndman's method is elegant and conceptually appealing, it in fact crucially depends on very effective but non-robust one-dimensional time series methodology applied to series of principal component scores (\emph{fts} R package) and nonrobust dispersion matrix of forecasts responsible for the forecasts reconciliation. Although one may easily robustify their approach using some robust alternatives for the nonrobust building block of their approach-we loose its mathematical elegance and significantly increase its computational complexity (which  is very high even without any modifications).  
\\ The robustness of our method to outliers does not heavily depend on the type of functional outliers, which seems to be surprising as we have expected, that it should be different for the functional shape outliers, functional amplitude outliers, and for functional outliers with respect to the covariance structure. We have conducted several simulations, but for the sake of presentation of our method, herein we show some simulations proceeded for the 100 observations of Gaussian process with mean equal to $\mu (t)=\sin(4\pi\cdot t)$ and covariance function $C(s,t)=\alpha \cdot {{e}^{\left( -\beta \left| s-t \right| \right)}}=0.2 \cdot {{e}^{\left( -0.8 \cdot\left| s-t \right| \right)}}.$
The contamination comes from outlying observations taken from the Gaussian process with mean equal to 
$\mu_C(t)=\sin(2\pi\cdot t+\pi/2)$ and covariance function equal to $C(s,t).$
\begin{figure*}
\includegraphics[width=\textwidth]
{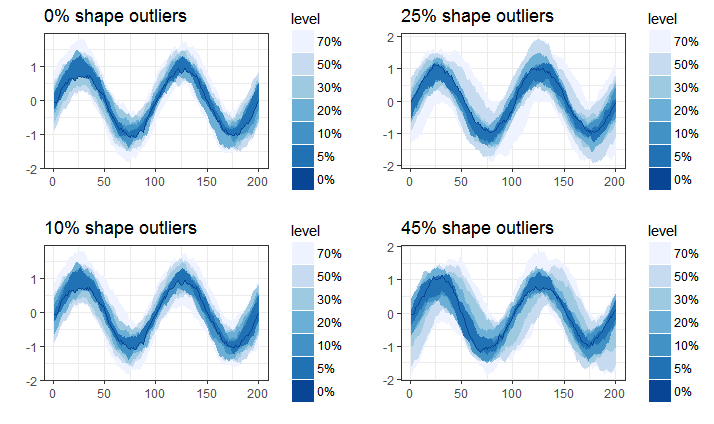}
\caption{Functional boxplot for data contaminated with 0\%, 10\%, 25\%, and 45\% shape outliers.}
\label{fig:1}
\end{figure*}
The robustness of the MBD median is visually evaluated with the following figures. The data has been contaminated with 0\%, 10\%, 25\%, and 45\% shape outliers. Functional boxplot for data contaminated with 0\%, 10\%, 25\%, and 45\% shape outliers is presented in Figure \ref{fig:1}.
\\ 
Data contaminated with 10\% and 45\% shape outlying observations as well as outliergrams for the data are presented in Figures \ref{fig:2} and \ref{fig:3}. 
It is worth to notice that the outliergram, that has been introduced by Arribas-Gil and Romo \cite{Arribas} to detect shape outliers,  
does not work well, when the number of outliers is about 45\%. The outliergram then stops detecting outliers with respect to shape.
Figures containing outliegram were prepared using a \emph{roahd} R-package (see \cite{Tarabelloni}) and functional boxplots were prepared using a \emph{DepthProc} R-package (see \cite{Kos1}) although it is straightforward to obtain outliergram in \emph{DepthProc} as well.
\\ Taking into account the fact that for a fixed $\alpha$, a volume of the $\alpha-$central region may be treated as a dispersion measure \cite{LPS}, comparing functional boxplots is a sufficient way to compare "effectiveness" (see Figure \ref{fig:5}). A comparison of functional time series predictor "effectiveness" may be conducted in terms of a comparison of volumes of $\alpha-$central regions or in terms of speeds of expansion of these regions as a functions of $\alpha$ (scale curve, see \cite{Lopez_scale}).
In our opinion, this fully nonparametric and moment-free data-analytic method taken from the multivariate case is the best solution in the functional case, where taking reasonable assumptions on the data-generating processes is often doubtful (there in no Lebesgue measure analogue in the $L^2[0,T]$ space).

\subsection{Computational complexivity} 

In order to compare a computational complexity of our proposal with Shang and Hyndman's proposal, we have considered empirical functional time series related to day and night monitoring of the Internet service (web portal) divided into four sub-services (for further details see Section \ref{sec:6}). The monitoring was conducted for 365 days of 2015. In other words, in/at the beginning we considered dataset consisted of five matrices, each of dimension $365\times 24$. For comparing two forecasting methods, we have considered forecasts obtained basing on moving window of length 10 observations.
A time of calculation of the forecasts using Shang and Hyndman's method was 37-38 min, whereas using the proposed aggregated median method was about 2-3 min. In both cases, we used the same software and hardware environment (WIN10,
mobile intel I7, 16GB RAM).
\\ Note that Shang and Hyndman's \cite{Shang} and Hyndman et al. \cite{Hyndman2011} indicated certain inconveniences of their methods related to an application of the generalized least squares in case of sparse design matrices. They listed certain theoretically interesting remedies for these inconveniences. In our opinion, the remedies are insufficient for the Internet data streams' analysis (for details see \cite{Kos2016}).
\section{Simulation study of our proposal}
A performance of FDA procedures in case of time dependent data is still an open and very intensively studied issue \cite{Hormann,Rana}.
In order to study the performance of our proposal in case of a setup departing from independent and identically distributed  case, we consider a hierarchical system consisted of relatively popular functional autoregression processes of order one (FAR(1)). 
\subsection{FAR(1) process}
Functional autoregression process of order one, i.e. FAR(1), is defined as
$$X_{n+1}=\Psi (X_n)+\epsilon_{n+1},$$
where both errors and the observations are functions,
and the operator
\\ $\Psi:L^2(\Omega)\to L^2(\Omega)$ is a linear operator transforming a function into another function, i.e., $\Psi(X)(t)=\int k(t,s)X(s)ds$, where $k(t,s$) is a bivariate kernel. 
Note that each kernel can be rewritten in the form (Mercer's theorem)
\begin{equation}
k(s,t)=\sum_{n=1}^{+\infty}\lambda_n e_n(t)e_n(s),
\end{equation}
where $\lambda_n$ are eigenvalues such that $\lambda_1>\lambda_2>...>0$ and $\{e_n\}_{n=1}^{+\infty}$ is an orthonormal basis in $L^2(\Omega)$ space. 
If the kernel is continuous, the series converges uniformly. Moreover, the population eigenfunctions
can be consistently estimated by the empirical eigenfunctions (see Theorem 13.2 of Horv\'ath and Kokoszka (2012) \cite{Horvath}). 
\\ The popular condition $||k|| < 1$ ensures the existence of a stationary solution to FAR(1) equations. 
However, the condition can be weakened as in Lemma 13.1 of Horv\'ath and Kokoszka (2012) \cite{Horvath} and in the simulation study we use a $C0$ condition of the Lemma: there exists a positive integer $j_0$ such that $||\Psi^{j_0}||<1$. The condition ensures  the existence of
a unique stationary causal solution to FAR(1) equations (see Theorem 13.1 of \cite{Horvath}).
\\ Moreover, H\"ormann and Kokoszka \cite{Hormann} showed (see their Example 2.1) that FAR(1) process with zero mean error is also $L^p-m-approximable$ with $p\geq 2$ (without requiring any smoothness properties for the innovations process), which is an analogue to the moment-based notion
of dependence in FTS setup. Thus, considering FAR(1) process we take into account the weak temporal dependence between the functional random variables. The considered example shows that our proposal for functional data is robust to some kind of temporal dependencies, i.e. it works for functional observations generated from FAR(1) process.
In the simulation framework a strength of a time-dependency between observations from the FTS sample may alternatively be taken into account using Kendall correlation coefficient for functional data (see \cite{kendal}).
\subsection{Hierarchical structure simulation framework}
\begin{figure*}
\includegraphics[width=0.5\linewidth]{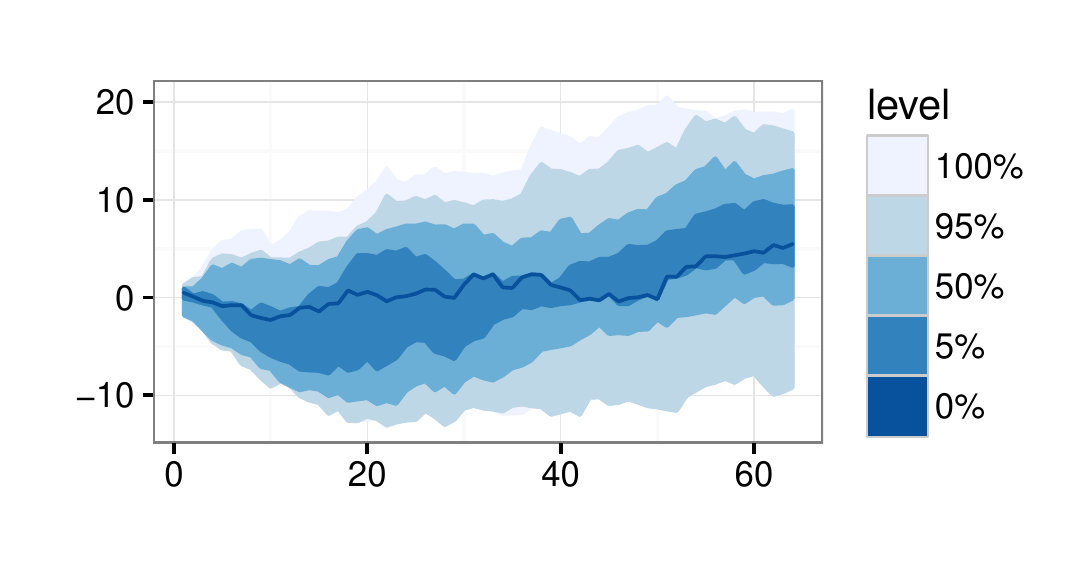}
\includegraphics[width=0.5\linewidth]{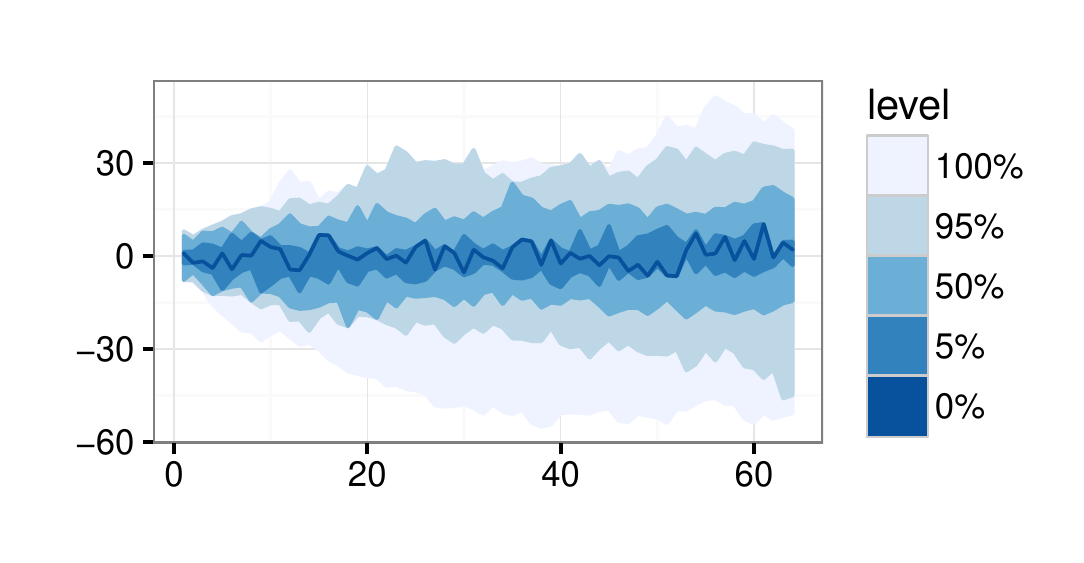}
\caption{Functional boxplots for prediction errors for the top level of the FAR(1) HFTS, without outliers (left) and with 10\% shape outliers (right). 
Moving mean method has been applied, window length $k=10$, \textit{DepthProc} R package.}
\label{fig:movingmean}
\end{figure*}

\begin{figure*}
\includegraphics[width=0.5\linewidth]{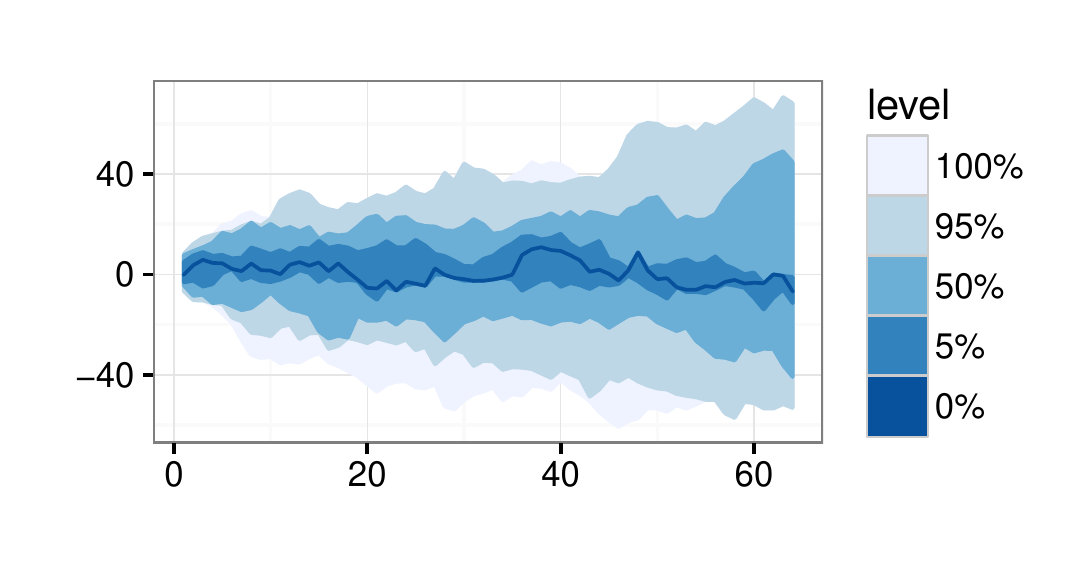}
\includegraphics[width=0.5\linewidth]{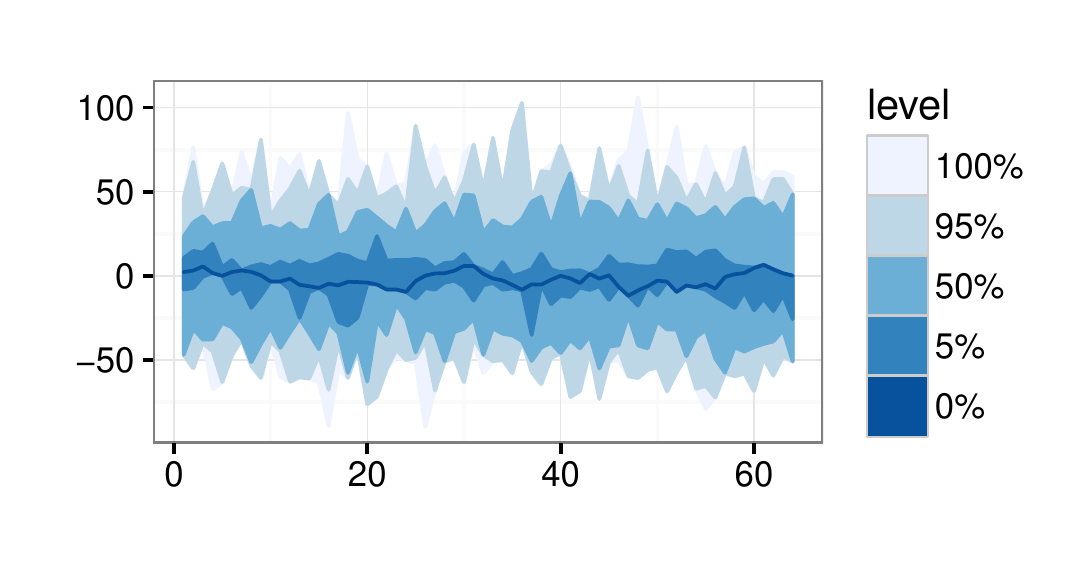}
\caption{Functional boxplots for prediction errors for the top level of the Wiener HFTS, without outliers (left) and with 10\% shape outliers (right). Aggregated functional median method has been applied, window length $k=10$, \textit{DepthProc} R package.}
\label{fig:wienerFM}
\end{figure*}

\begin{figure*}
\includegraphics[width=0.5\linewidth]{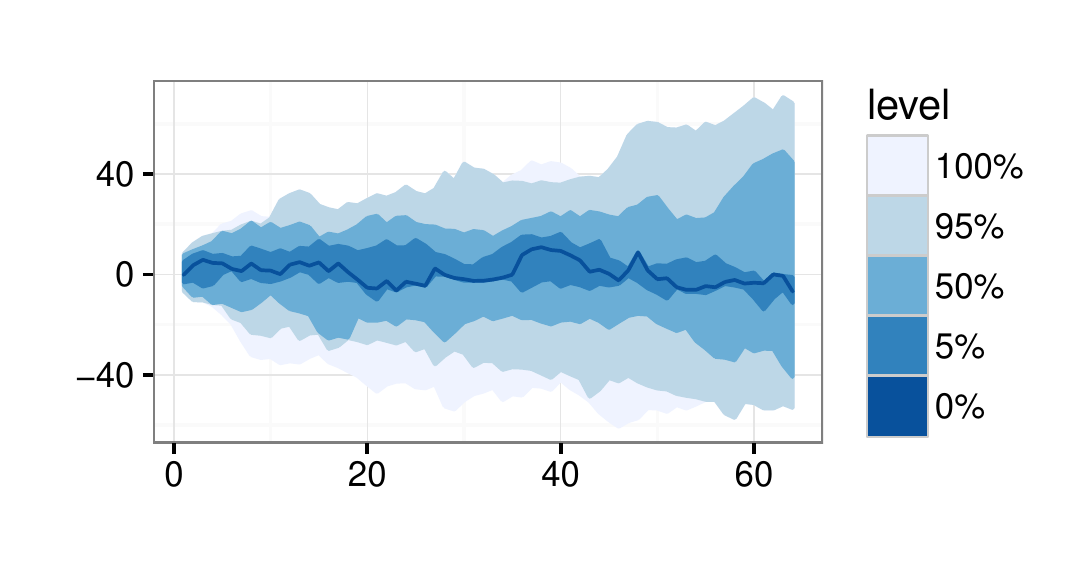}
\includegraphics[width=0.5\linewidth]{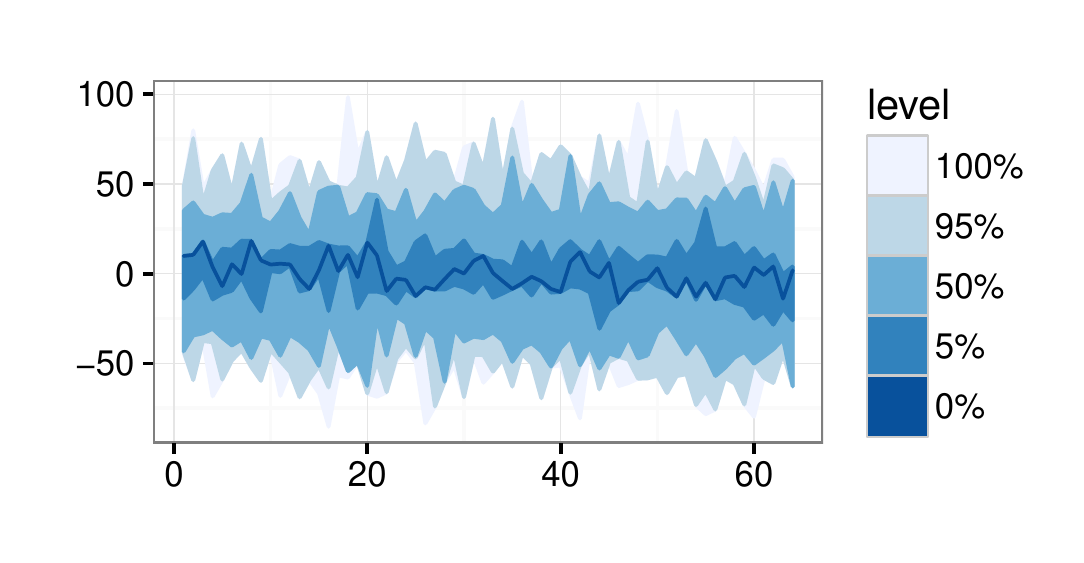}
\caption{Functional boxplots for prediction errors for the top level of the Wiener HFTS, without outliers (left) and with 10\% shape outliers (right). 
Moving mean method has been applied, window length $k=10$, \textit{DepthProc} R package.}
\label{fig:wienerMM}
\end{figure*}

Consider a hierarchical structure depicted in Figure \ref{fig_hier}, where each process at the bottom level (in each node at the lowest level of the hierarchy) is a FAR(1) process and the process at higher level is a sum of processes at the lower level below him plus an error, for example, $H^1_2=H^1_3+H^2_3+H^3_3+e^1_2$,
$H_1=H^1_2+H^2_2+H^3_2+e_1$, and $H=H_1+H_2+e$, and all errors, namely $e^1_2,e^2_2,e^3_2,e^4_2,e^5_2,e^6_2,e_1,e_2,e$, are mutually independent standard Wiener processes multiplied by $\frac{1}{10}$, for example, $e=e(t)=\frac1{10}W(t)$, so that errors do not dominate the FAR(1) processes.
The aim of the simulation study is to obtain a forecast with our proposal jointly at all levels of the hierarchy.
\\ To sum up, we give some details about the simulation study. The FAR(1) data generating process series in each node at the bottom level are generated in the simulation according to the model
$$X_{n+1}(t)=\int_0^Tk(s,t)X_n(s)ds+\epsilon_{n+1}(t),$$ where $n = 0,...,N-1$ and error $\epsilon_{n+1}(t)$ is a standard Wiener process. 
\\ Following \cite{Dider} we decided to consider kernels of the following form:
\\ sloping plane of $s$, i.e. $k(s,t)=Cs$, 
\\ sloping plane of $t$, i.e. $k(s,t)=Ct$,
\\ and exponential kernel, i.e. $k(s,t)=C\exp\{-\frac{|s-t|}{2}\}$. However, for the clarity of the simulation study presentation, we show only the simulations performed with the most complicated exponential kernel.
\\ The normalizing constants $C$ are chosen so that $||\Psi||=||k||=0.5$ or \\ $||\Psi||=||k||=0.8$.
\\
\\ In the HFTS simulation studies we used the following software and hardware environment: I7 6700, Ubuntu 16.04 LTS, 128 GB RAM. 
\\ We used a free R package $far$ (see \cite{far}) to generate samples from the FAR(1) process for nodes at the bottom level of the considered hierarchy.
We used a \textit{simul.far.wiener} command with a parameter \textit{d.rho}, which is an approximation of the operator $\Psi$ in Karhunen-Lo\'eve basis truncated to three eigenfunctions.
We generated 100 FAR(1) samples for each of the 18 nodes at the bottom level of the hierarchy. We calculated a moving functional median w.r.t. MBD and GBD with a window length equal to 3, 5 and 10 observations for each node at the bottom level of the hierarchy. Afterwards, the functional samples for the three corresponding nodes at the bottom level of the hierarchy and corresponding error $e_2^i$ have been added and thus a six perturbated samples, namely $X^{H_2^i}_{n}(t)$, where $1=1,2,\dots,6$, for nodes in the lowest but one level of the hierarchy have been obtained (see Figure \ref{fig_hier}). Simultaneously, six forecasts for nodes $H_2^1,\dots,H_2^6$ have been calculated, for example, the forecast for $H_2^1$ is $$\hat{X}_{n+1}^{H_2^1}(t)=MED_{FD}\{X^{H_3^1}_{n}(t),\dots,X^{H_3^1}_{n-k+1}(t)\}+$$
$$+MED_{FD}\{X^{H_3^2}_{n}(t),\dots,X^{H_3^2}_{n-k+1}(t)\}+MED_{FD}\{X^{H_3^3}_{n}(t),\dots,X^{H_3^3}_{n-k+1}(t)\},$$
where $FD$ denotes chosen functional depth, $GBD$ or $MBD$, and $k$ is a length of the moving window, i.e. $k=3,5,10$. 
\\ The functional samples from nodes $H_2^i$, $i=1,2,3$ and the corresponding error $e_1$ and functional samples from nodes $H_2^i$, $i=4,5,6$ and the corresponding error $e_2$ have been added and two perturbated samples for nodes in the top but one level of the hierarchy have been obtained, namely $X^{H_1}_{n}(t)$ and $X^{H_2}_{n}(t)$.
Simultaneously, two forecasts have been calculated, for example, the forecast for $H^1$ is 
$$\hat{X}_{n+1}^{H^1}(t)=MED_{FD}\{X^{H_2^1}_{n}(t),\dots,X^{H_2^1}_{n-k+1}(t)\}+$$
$$+MED_{FD}\{X^{H_2^2}_{n}(t),\dots,X^{H_2^2}_{n-k+1}(t)\}+MED_{FD}\{X^{H_2^3}_{n}(t),\dots,X^{H_2^3}_{n-k+1}(t)\}.$$
\\ Finally, the two functional samples from nodes $H_1$ and $H_2$ and the error $e$ have been added and we obtain a perturbated sample for a node in the top level of the hierarchy, i.e. $X^H_n(t)$. Forecast at the node in the top level of the hierarchy is given by the formula
$$\hat{X}_{n+1}^{H}(t)=MED_{FD}\{X^{H_1}_{n}(t),\dots,X^{H_1}_{n-k+1}(t)\}+$$
$$+MED_{FD}\{X^{H_2}_{n}(t),\dots,X^{H_2}_{n-k+1}(t)\}.$$
In this way, we have obtained a functional hierarchical structure and a forecast for each node in the considered hierarchical structure. We repeated the whole experiment for thirty times.
\\ \emph{Note} that because of the shapes of functions generated from FAR(1) process, functional medians obtained with $GBD$ and $MBD$ are very similar. Thus, it is enough to consider functional medians induced by $MBD$. Note also that in economics it is especially important to monitor the persistence of certain property for the longest possible uninterrupted subinterval. That is why from a merit point of view it is worth to discriminate between $GBD$ and $MBD$ because of the fact that when using $GBD$ we focus our attention on a certain part of the functions domain. 
\\ The quality of the forecast has been evaluated with the mean absolute forecast error (MAFE, see \cite{Shang}) for each of the four levels of the hierarchy (see Tables \ref{tab:far}-\ref{tab:out40}), and functional boxplots have been made for each of 27 nodes of the considered hierarchical structure (see Figures \ref{fig:hfts_comp}-\ref{fig:wienerMM}). 
\\ Functional boxplots and volumes of the $\alpha-$central regions, in particular, have been used in order to evaluate "effectiveness" of the predictors. We have compared forecasts obtained with our method, with a moving functional mean method, for contaminated, and for clean datasets (for example, see Figures \ref{fig:hfts_comp} and \ref{fig:movingmean} for a comparison on the top level).
\\ In a case of the contaminated data we replaced some fraction of observations in the FAR(1) samples in nodes on the bottom level of the hierarchy with 5\%, 10\% and 40\% outlying functions.
The contamination data have been generated from the following process 
$f_s(t)=60W_1(t)\sin(2\pi t)+\sqrt{2}W_2(t)\cos(2\pi t)$, where $W_1(t)$ and $W_2(t)$ are standard Wiener processes.
\\The contamination involved both \emph{shape outliers} as well as \emph{magnitude outliers}, that have been generated from the process $f_s(t)$. 
The outliergram of Arribas-Gil and Romo \cite{Arribas} has been used to verify whether the replaced functions are actually shape outliers, and adjusted functional boxplot of Tarabelloni \cite{Tarabelloni} has been used to verify whether the replaced functions are actually functional magnitude outliers.
The outlying functions have been placed among the observations at the nodes at the bottom level of the hierarchy at random using binomial distribution where parameter $p$ (the probability of success) is a fraction of outlying functions.
Finally, we used our aggregated median method to obtain a forecast for all nodes of the hierarchy. 
\\ The robustness of our method has been evaluated in terms of changes in MAFEs and functional boxplots for contaminated samples in a comparison to clean samples.
From Tables \ref{tab:far} - \ref{tab:out40} one may conclude that our method allows for a relatively robust forecasting with respect to both shape and magnitude outliers in case of weak dependence between functional observations. Conducting an extensive simulation study involved a comparison of functional boxplots for each out of 27 nodes of the considered hierarchy for our method and for moving functional mean method (see Fig. \ref{fig:hfts_comp} and Fig. \ref{fig:movingmean} for a comparison on the top level).
Similarly, from Tables \ref{tab:far}-\ref{tab:out40} one may conclude, that our method is relatively robust for identically distributed functional data.
\\ In Figure \ref{fig:hfts_comp} there are examples of functional boxplots for prediction errors, which are functions being differences between the true value and the forecast obtained with an aggregated functional median method with a moving window of length $10$, for the top level of the FAR(1) HFTS, without outliers (left) and with 10\% shape outliers (right). In Figure \ref{fig:movingmean} there are examples of functional boxplots for prediction errors, which are functions being differences between the true value and the forecast obtained with a moving functional mean method with a moving window of length $10$, for the top level of the FAR(1) HFTS, without outliers (left) and with 10\% shape outliers (right). 
\\
\\ \textit{Brief summary of HFTS simulation study results in the case of FAR(1):}
\\
\\ 1. MAFE increases with the level of the hierarchy.
\\ 2. MAFE increases with an amount of the contamination for both methods for any level, as well.
\\ 3. In the case of clean data, the longer is the moving window, the smaller is MAFE for aggregated functional median and for moving mean methods.
\\ 4. The longer is the moving window the worse in terms of MAFE is an aggregated median forecast in comparison to a moving functional mean forecast (see Table \ref{tab:far}). The aggregated functional median outperforms the moving functional mean method in case of contaminated data. However, in case of clean data a situation is unclear and the moving mean method often outperforms our method. 
\\ 5. It can be observed that while the central regions of the functional boxplots are expanding with a domain of the function, the aggregated functional median is close to zero function (see Figures \ref{fig:hfts_comp} and \ref{fig:movingmean}), which suggest a median-unbiasedness. 
\\ 6. Our procedure seems to be robust, i.e. even 10\% of outlying functions do not significantly distort functional medians and volumes of the central regions of the functional boxplots - it is discernible at the functional boxplots (see Figures \ref{fig:hfts_comp} and \ref{fig:movingmean}). 
\begin{table}
\caption{MAFE for moving functional mean and aggregated median methods, where FAR(1) sample is not contaminated with outliers.}
\label{tab:far} 
\begin{center}
\begin{small}
\begin{tabular}{c|ccc|cccc}\hline
\noalign{\smallskip}
 & &Moving mean&  &  &Aggregated median&  \\
	$k$-window length& $3$ & $5$ & $10$ & $3$ &$5$ & $10$ \\\hline
	Bottom level & 0.69 & 0.68 &0.65&0.67 & 0.69&0.70\\
	Level 2 &1.29& 0.96 &0.77 & 1.47 & 1.01 & 0.90  \\
	Level 3 &3.88  &1.77&1.31 & 4.41 & 1.68 & 1.50 \\
    Top level & 11.65 & 2.43 & 1.85& 8.83 & 2.43 & 2.02   \\
\hline	
\end{tabular}
\end{small}
\end{center}
\end{table}
\begin{table}
\caption{MAFE for moving functional mean and aggregated median methods, where FAR(1) sample is contaminated with 10\% outliers, which are both shape and magnitude outliers. The outliers are generated from $f_s(t)$ process.}
\label{tab:out10} 
\begin{center}
\begin{small}
\begin{tabular}{c|ccc|cccc}\hline
\noalign{\smallskip}
 & &Moving mean&  &  &Aggregated median&  \\
	$k$-window length& $3$ & $5$ & $10$ & $3$ &$5$ & $10$ \\\hline
	Bottom level & 2.41 & 2.29 &2.12&1.70 &1.69&1.71\\
	Level 2 &5.80&  5.41 &5.04 & 4,61& 4,59&4,59   \\
	Level 3 &12.01  &11.55& 11.06 & 10.74&  11.33&10.72  \\
    Top level & 17.53 & 17.92 & 29.54 & 17.94& 17.47 &17.41   \\
\hline	
\end{tabular}
\end{small}
\end{center}
\end{table}
\begin{table}
\caption{MAFE for moving functional mean and aggregated median methods, where FAR(1) sample is contaminated with 40\% outliers, which are both shape and magnitude outliers. The outliers are generated from $f_s(t)$ process.}
\label{tab:out40} 
\begin{center}
\begin{small}
\begin{tabular}{c|ccc|cccc}\hline
\noalign{\smallskip}
 & &Moving mean&  &  &Aggregated median&  \\
	$k$-window length& $3$ & $5$ & $10$ & $3$ &$5$ & $10$ \\\hline
	Bottom level & 5.65 & 5.25 & 4.84 & 4.51&4.10 & 4.11 \\
	Level 2 &11.00&  10.31 &9.83 & 9.87 & 9.13 & 9.17  \\
	Level 3 &20.24  &19.28& 18.72 & 19.15 & 17.99 & 18.08 \\
    Top level & 27.82 & 27.91 & 28.43 & 28.13 & 26.63 & 26.88 \\
\hline	
\end{tabular}
\end{small}
\end{center}
\end{table}
\begin{figure}
\ovalbox{
\begin{tikzpicture}[auto, level 1/.style={sibling distance=57mm}, level 2/.style={sibling distance=20mm}, level 3/.style={sibling distance=5mm}]
\node {$H$}
  child {node {$H_{1}$}
    child {node {$H^{1}_{2}$}
     child {node {$H_3^{1}$}}
     child {node {$H_3^{2}$}}
      child {node {$H_3^{3}$}}
      }
    child {node {$H^{2}_2$}
      child {node {$H_3^{4}$}}
      child {node {$H_3^{5}$}}
      child {node {$H_3^{6}$}}
    }
     child {node {$H_2^{3}$}
     child {node {$H_3^{7}$}}
     child {node {$H_3^{8}$}}
     child {node {$H_3^{9}$}}
    }
  }
  child {node {$H_{2}$}
  child {node {$H^{4}_{2}$}
   child {node {$H_3^{10}$}}
     child {node {$H_3^{11}$}}
    child {node {$H_3^{12}$}}
  }
    child {node {$H^{5}_{2}$}
      child {node {$H_3^{13}$}}
      child {node {$H_3^{14}$}}
      child {node {$H_3^{15}$}}
    }
  child {node {$H^{6}_{2}$}
    child {node {$H_3^{16}$}}
     child {node {$H_3^{17}$}}
    child {node {$H_3^{18}$}}
  }
};
\end{tikzpicture}}
\caption{Hierarchical structure of considered FAR(1) process}
\label{fig_hier}
\end{figure}
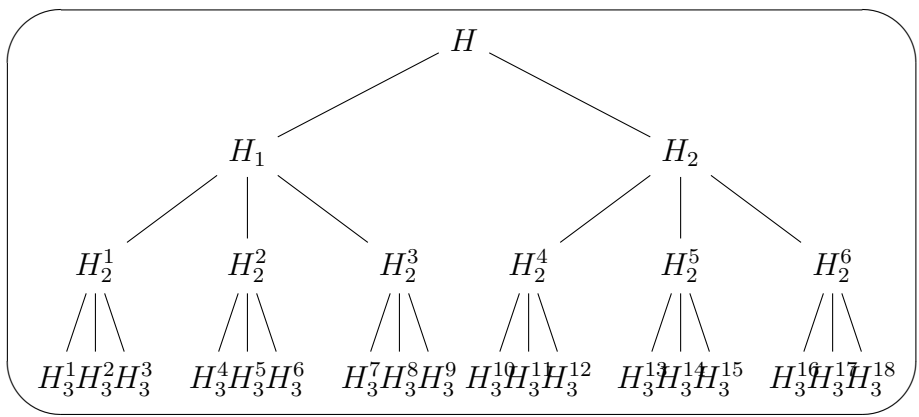
\begin{figure*}
\includegraphics[width=0.5\linewidth]{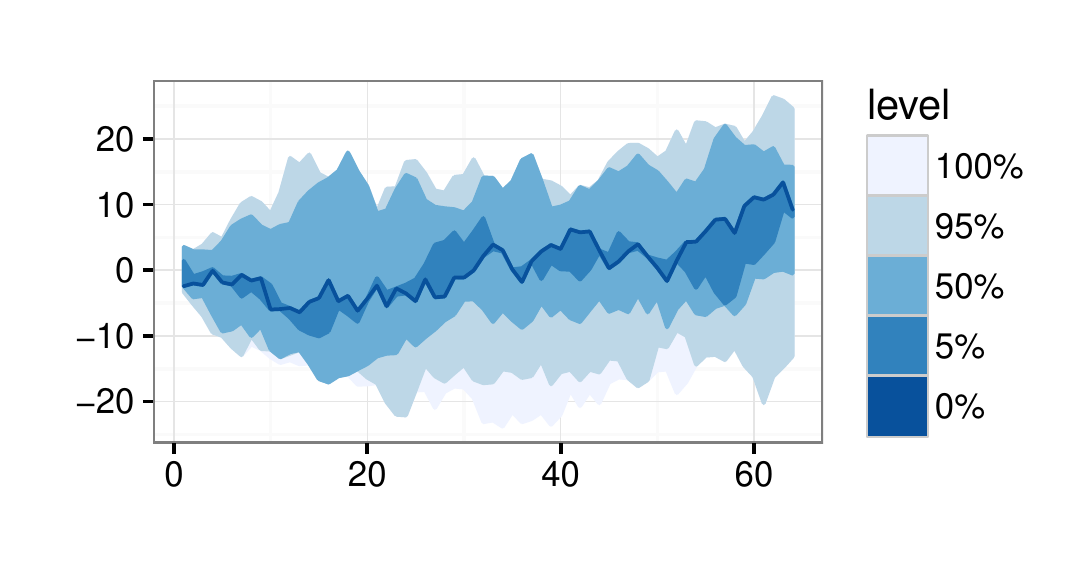}
\includegraphics[width=0.5\linewidth]{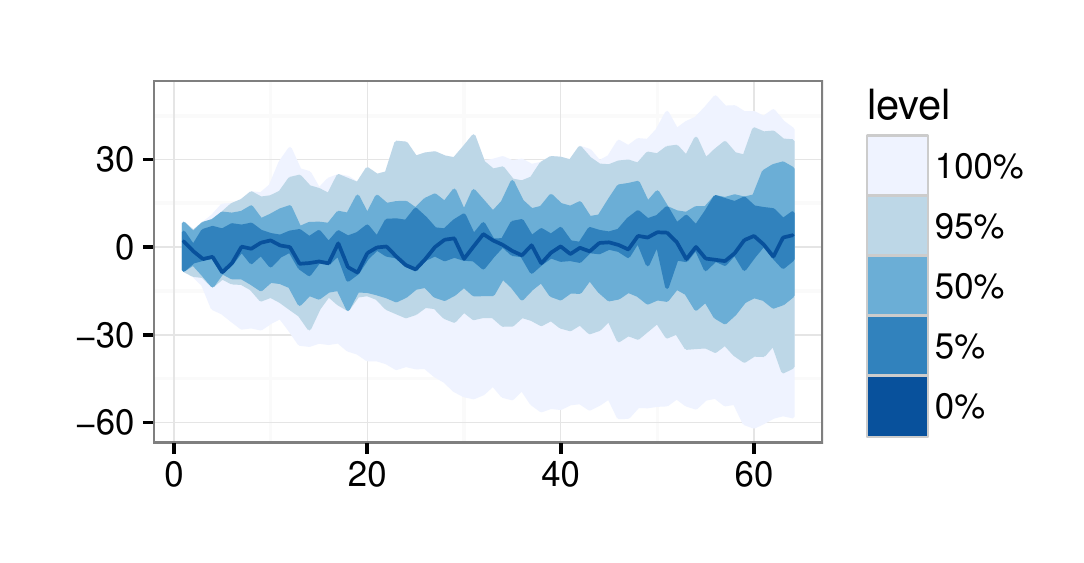}
\caption{Functional boxplots for prediction errors for the top level of the FAR(1) HFTS, without outliers (left) and with 10\% shape outliers (right). Aggregated functional median method has been applied, window length $k=10$, \textit{DepthProc} R package.}
\label{fig:hfts_comp}
\end{figure*}
\\
\\ We considered a hierarchical structure depicted in Figure \ref{fig_hier}, where each process at the bottom level (in each node at the lowest level of the hierarchy) is a standard Wiener process times $10$ and the process at higher level is a sum of processes at the lower level below it plus an error, just as described above for the FAR(1) case. In other words, the only difference with a situation described above is that we consider $10\times W(t)$ process in each node at the lowest level of the hierarchy instead of FAR(1) process. 
The robustness of our method has been evaluated in terms of changes in MAFE for contaminated samples in a comparison to clean samples (see Tables \ref{tab:wiener}-\ref{tab:wiener40}).
\\ In Figure \ref{fig:wienerFM} there are examples of functional boxplots for prediction errors, which are functions being differences between the true value and the forecast obtained with an aggregated functional median method with a moving window of length $10$, for the top level of the Wiener HFTS, without outliers (left) and with 10\% shape outliers (right). In Figure \ref{fig:wienerMM} there are examples of functional boxplots for prediction errors, which are functions being differences between the true value and the forecast obtained with a moving functional mean method with a moving window of length $10$, for the top level of the Wiener HFTS, without outliers (left) and with 10\% shape outliers (right). 
\\
\\ \textit{Brief summary of HFTS simulation study results in the case of Wiener process:}
\\ 1. MAFE increases with the level of the hierarchy. 
\\ 2. MAFE increases with an amount of the contamination for both methods for any level, as well.
\\ 3. In the case of clean data, the longer is the moving window, the smaller is MAFE for aggregated functional median and for moving mean methods. Both methods give comparable results (see Tables \ref{tab:wiener}-\ref{tab:wiener40}). 
\\ 4. The aggregated functional median outperforms the moving functional mean method in case of contaminated data.
\\ 5. It can be observed that while the central regions of the functional boxplots are expanding with a domain of the function, the aggregated functional median is close to zero function (see Figures \ref{fig:wienerFM} and \ref{fig:wienerMM}). 
\\ 6. Our procedure exhibits a robustness, i.e. even 10\% of outlying functions do not distort functional median and the central regions of the functional boxplots - it is discernible at functional boxplots (see Figures \ref{fig:wienerFM} and \ref{fig:wienerMM}).  

\begin{table}
\caption{MAFE for moving functional mean and aggregated median methods, where Wiener process sample is not contaminated with outliers.}
\label{tab:wiener} 
\begin{center}
\begin{small}
\begin{tabular}{c|ccc|cccc}\hline
\noalign{\smallskip}
 & &Moving mean&  &  &Aggregated median&  \\
	$k$-window length& $3$ & $5$ & $10$ & $3$ &$5$ & $10$ \\\hline
	Bottom level & 0.72
 & 0.68
 &0.65
&0.74
 & 0.69
&0.70
\\
	Level 2 &1.95
& 0.96
 & 0.77
& 2.21
 & 1.01
 & 0.90
 \\
	Level 3 &5.86
  &1.77
& 1.31
& 6.63
 & 1.68
&  1.50
\\
    Top level & 11.72
 & 2.43
 & 1.85
& 13.27
 & 2.43
 &  2.02
 \\
\hline	
\end{tabular}
\end{small}
\end{center}
\end{table}
\begin{table}
\caption{MAFE for moving functional mean and aggregated median methods, where Wiener process sample is contaminated with 10\% outliers, which are both shape and magnitude outliers. The outliers are generated from $f_s(t)$ process.}
\label{tab:wiener10} 
\begin{center}
\begin{small}
\begin{tabular}{c|ccc|cccc}\hline
\noalign{\smallskip}
 & &Moving mean&  &  &Aggregated median&  \\
	$k$-window length& $3$ & $5$ & $10$ & $3$ &$5$ & $10$ \\\hline
	Bottom level &  2.80

& 0.68
 & 0.68
 &2.01 & 0.72
 &  0.73
\\
	Level 2 &6.29
&  0.93
 & 0.78
&5.15
 & 1.04
& 0.91
  \\
	Level 3 & 13.18
 &1.50
& 1.25
 &12.03
 & 1.70
 &  1.47
\\
    Top level & 19.12
 & 2.19
 &  1.77
& 19.56
& 2.45
 &  2.05
 \\
\hline	
\end{tabular}
\end{small}
\end{center}
\end{table}
\begin{table}
\caption{MAFE for moving functional mean and aggregated median methods, where Wiener process sample is contaminated with 40\% outliers, which are both shape and magnitude outliers. The outliers are generated from $f_s(t)$ process.}
\label{tab:wiener40} 
\begin{center}
\begin{small}
\begin{tabular}{c|ccc|cccc}\hline
\noalign{\smallskip}
 & &Moving mean&  &  &Aggregated median&  \\
	$k$-window length& $3$ & $5$ & $10$ & $3$ &$5$ & $10$ \\\hline
	Bottom level & 5.72 & 5.62
 & 5.13 & 4.57& 4.49
&  4.36

\\
	Level 2 &11.15& 10.86
  & 10.26
& 10.05 &  9.80
&   9.59
\\
	Level 3 &20.63  &20.43& 19.58
& 19.50 & 19.19
 &  18.85
\\
    Top level & 28.40 & 29.76
 &  28.67
& 28.69 & 28.46
&  28.07
\\
\hline	
\end{tabular}
\end{small}
\end{center}
\end{table}

\section{Empirical study}
\label{sec:6}
\begin{figure*}
\includegraphics[width=0.5\linewidth]{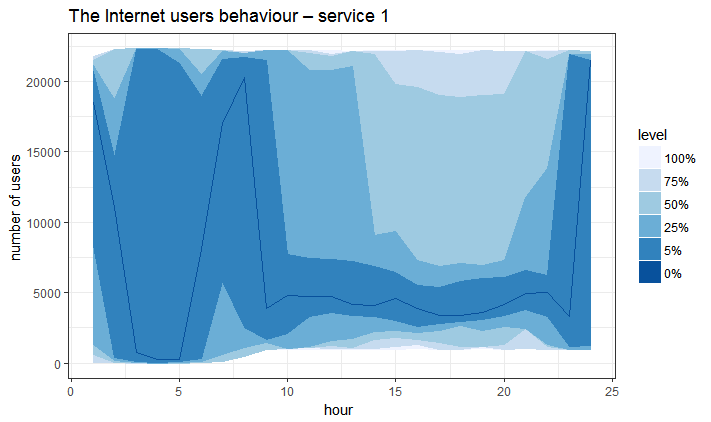}
\includegraphics[width=0.5\linewidth]{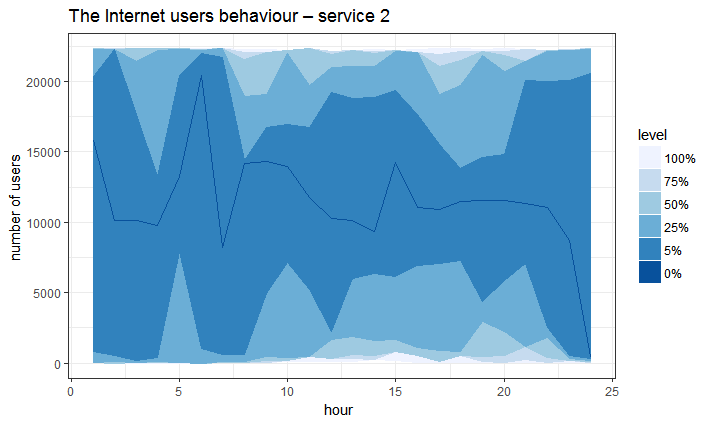}
\includegraphics[width=0.5\linewidth]{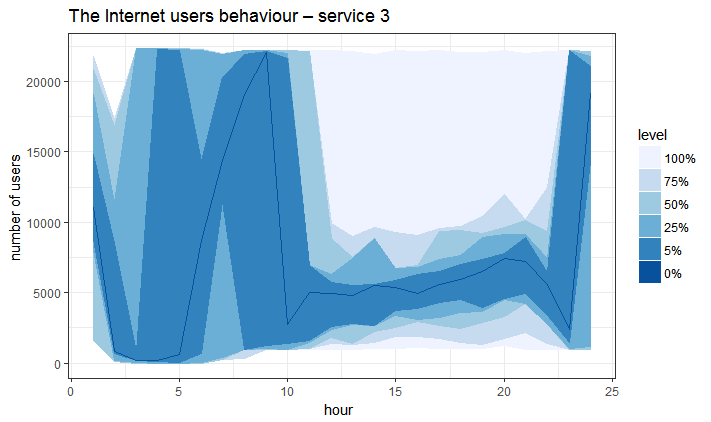}
\includegraphics[width=0.5\linewidth]{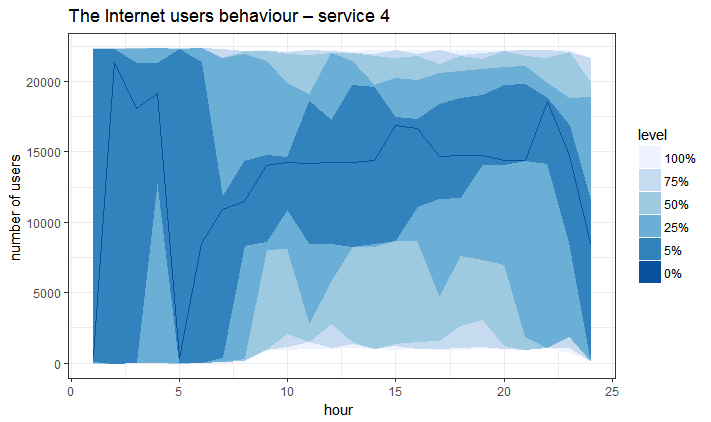}
\caption{Functional boxplots of number of four sub-services users, \textit{DepthProc} R package.}
\label{fig:5}
\end{figure*}

In order to show an empirical usefulness of our proposal, we have analyzed real dataset containing information on day and night monitoring of the Internet users' behaviors in a certain well-known Polish Internet service (web portal). The web portal is divided into four sub-services. We have received the data containing among others hourly number of users of the whole service as well as for four of its sub-services, called service 1 - 4 hereafter. The data come from 365 days of the year 2015. 
We treat the observations as continuous functions. 
As the statistical description of the considered dataset one may effectively use functional boxplots presented in Figure \ref{fig:5} (see \cite{SunGenton,Kos1}).

\begin{figure*}
\includegraphics[width=1\linewidth]{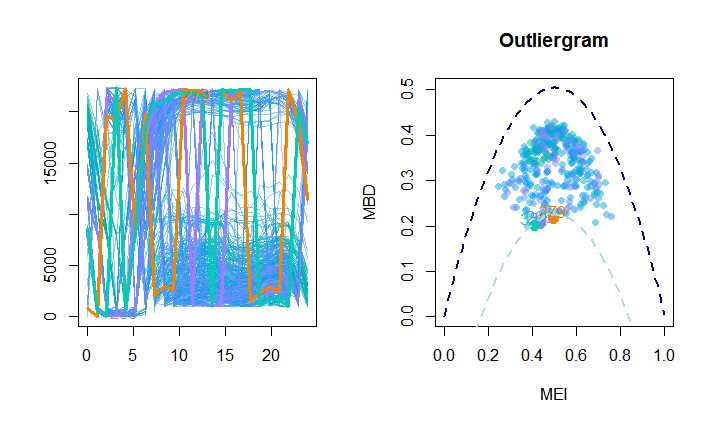}
\caption{Shape outliers and an outliergram for  service 1, \textit{roahd} R package.}
\label{fig:8}
\end{figure*}

\begin{figure*}
\includegraphics[width=1\linewidth]{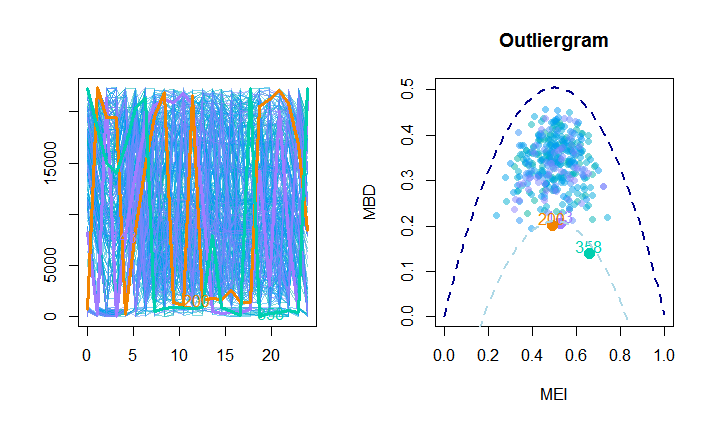}
\caption{Shape outliers and an outliergram for  service 2, \textit{roahd} R package.}
\label{fig:9}
\end{figure*}

\begin{figure*}
\includegraphics[width=1\linewidth]{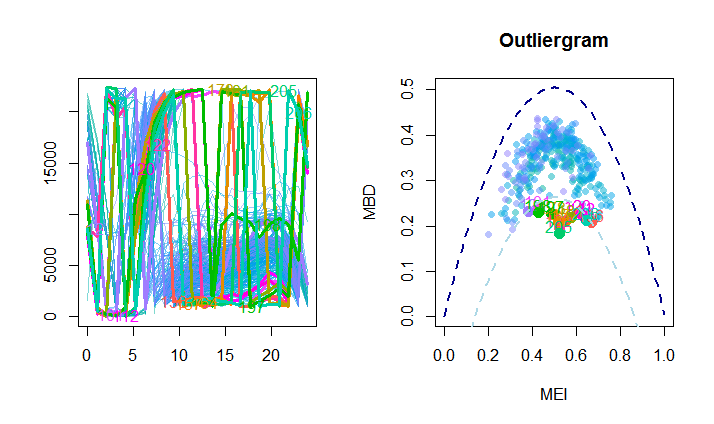}
\caption{Shape outliers and an outliergram for  service 3, \textit{roahd} R package.}
\label{fig:10}
\end{figure*}

\begin{figure*}
\includegraphics[width=1\linewidth]{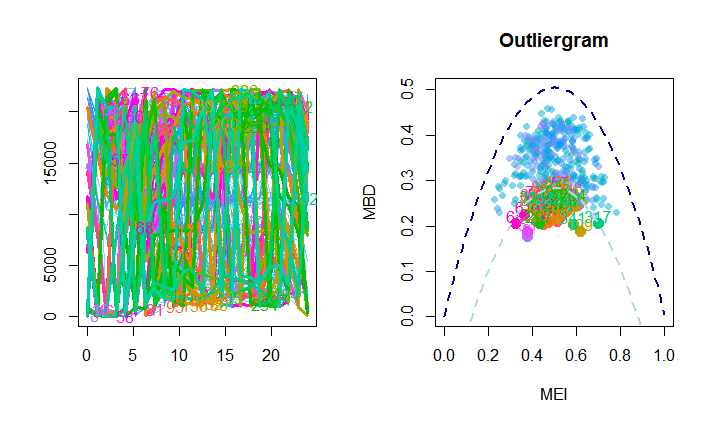}
\caption{Shape outliers and an outliergram for  service 4, \textit{roahd} R package.}
\label{fig:11}
\end{figure*}

\begin{figure*}
\includegraphics[width=1\linewidth]{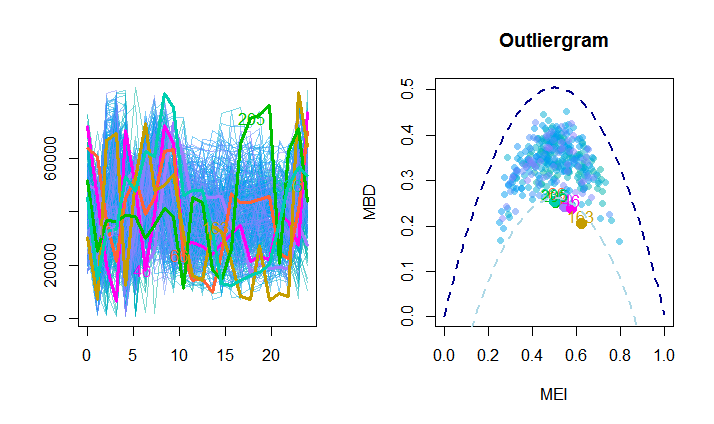}
\caption{Functional shape outliers and an outliergram for the whole analyzed service, \textit{roahd} R package.}
\label{fig:12}
\end{figure*}

\begin{figure*}
\includegraphics[width=\textwidth]{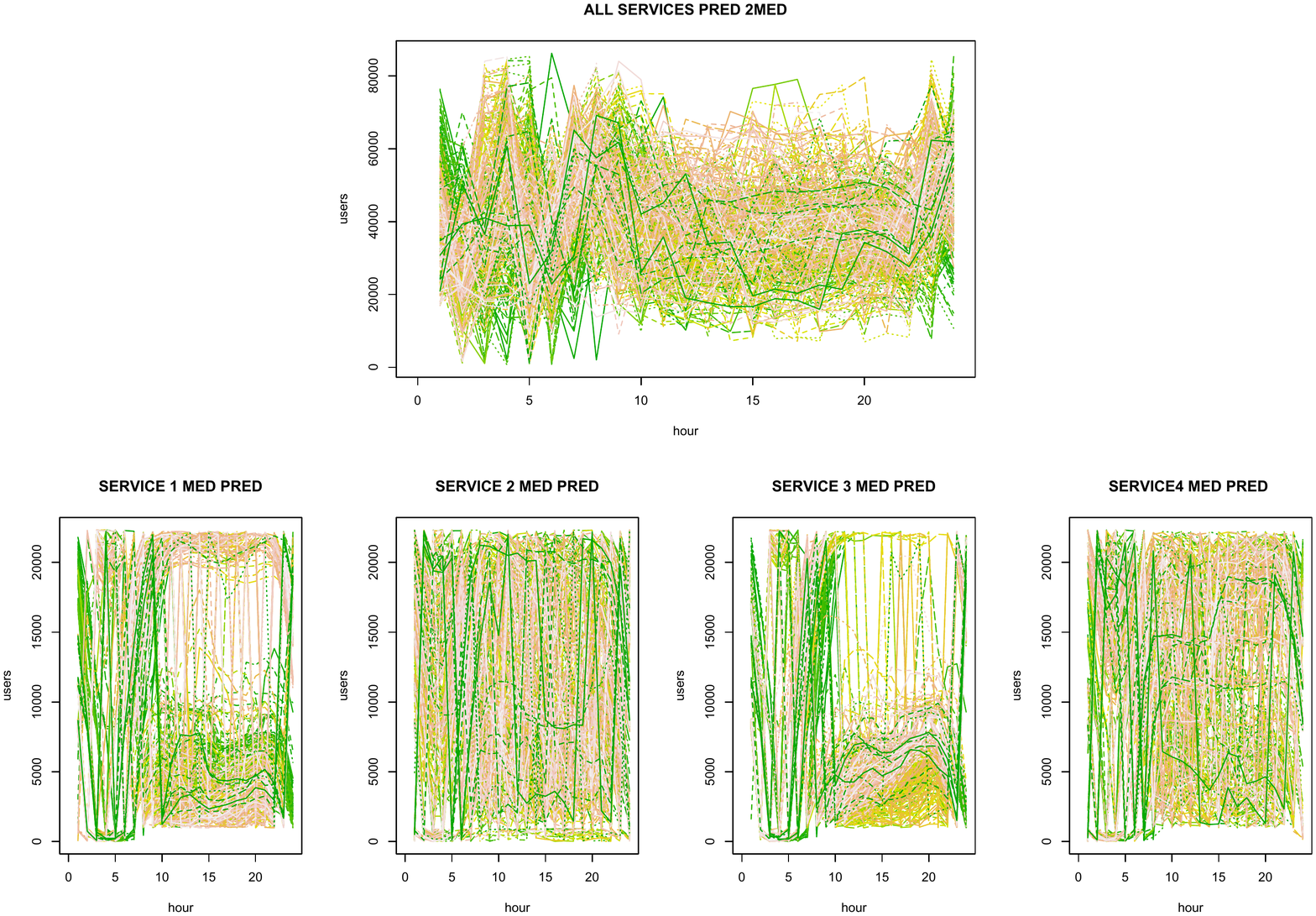}
\caption{The aggregated median predictor of the Internet services' users.}
\label{fig:13}
\end{figure*}

\begin{figure*}
\includegraphics[width=\textwidth]{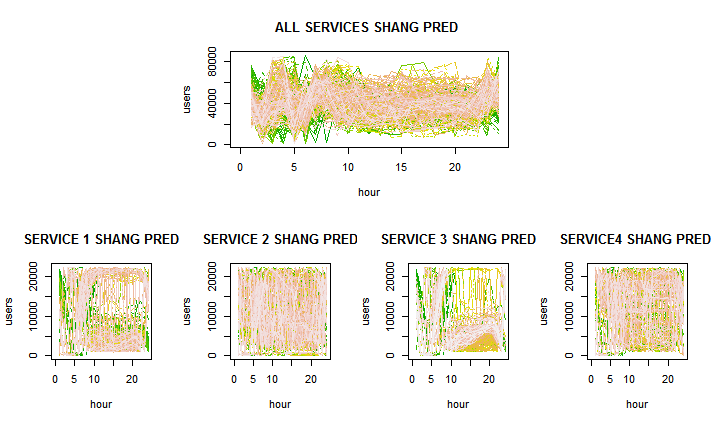}
\caption{The Shang and Hyndman predictor of the Internet services' users.}
\label{fig:14}
\end{figure*}

\begin{figure*}
\includegraphics[width=\textwidth]{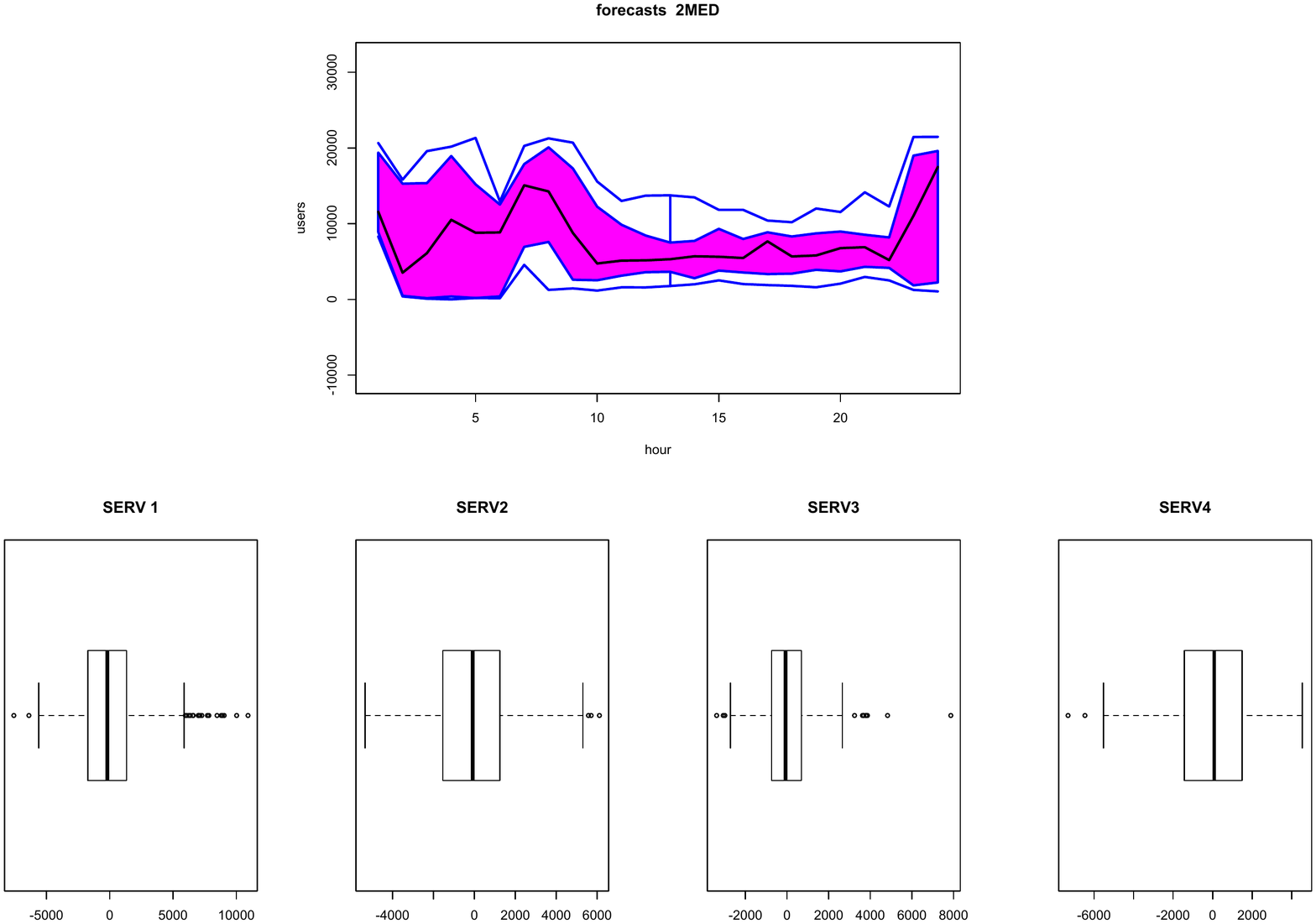}
\caption{
Four boxplots for the hourly average sum of the differences between the observed curves and the curves forecasted with the double functional median method. Above a functional boxplot for the forecasts of the number of service users, \textit{fda} R package.}
\label{fig:15}
\end{figure*}

\begin{figure*}
\includegraphics[width=\textwidth]{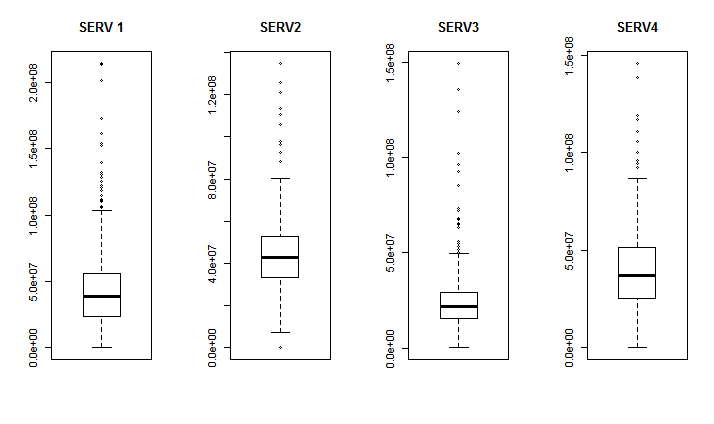}
\caption{
Four boxplots for the median of sum of squares of the differences between the observed curves and the curves forecasted with the aggregated functional median method.}
\label{fig:16}
\end{figure*}

\begin{figure*}
\includegraphics[width=\textwidth]{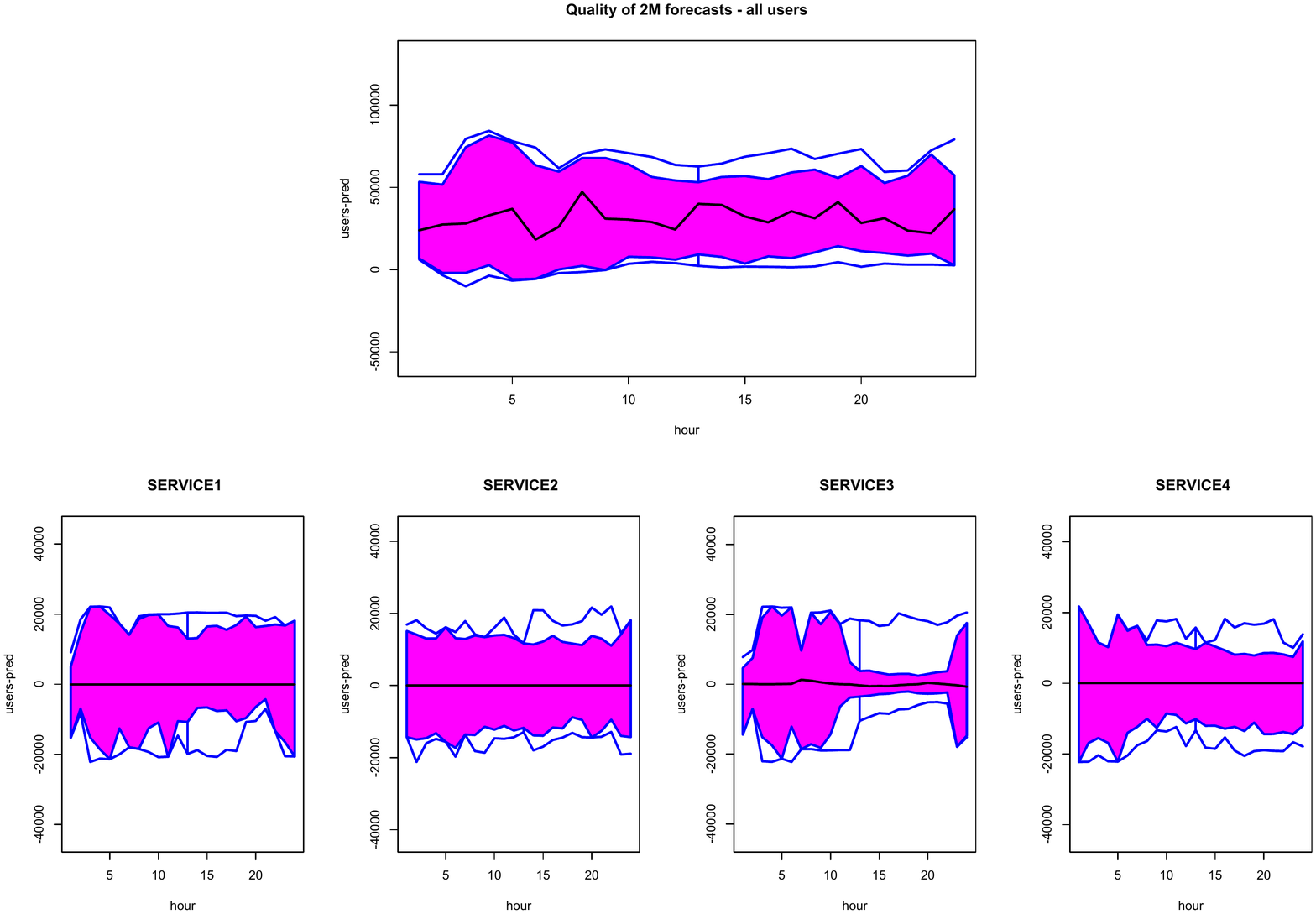}
\caption{
Five functional boxplots for the values of differences between the observed curves and the curves forecasted for four subservices users forecasted  with the aggregated functional median method, \textit{fda} R package.}
\label{fig:17}
\end{figure*}

\begin{figure*}
\includegraphics[width=\textwidth]{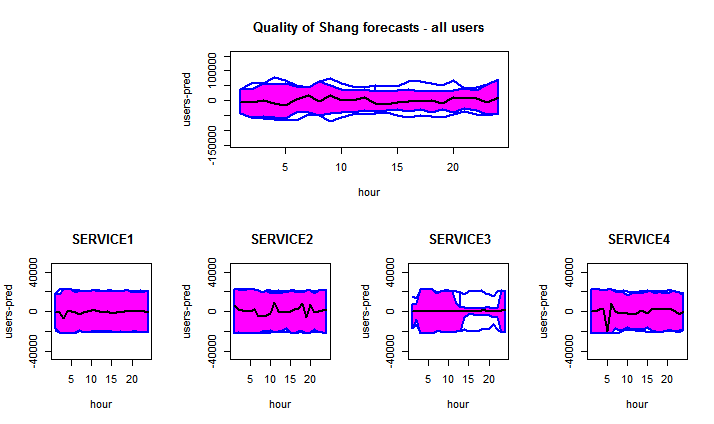}
\caption{
Five functional boxplots for the values of differences between the observed curves and the curves forecasted for four subservices users forecasted  with the Shang and Hyndman's method, \textit{fda} R package.}
\label{fig:18}
\end{figure*}

 The preliminary analysis of the Internet users' data included a procedure of detecting the functional magnitude outliers (see Figure \ref{fig:5}), and preparing outliergrams in order to detect the functional shape outliers (see Figures \ref{fig:8}-\ref{fig:11}) for the sub-services 1-4 and for the whole service (see Figure \ref{fig:12}). For definition and interpretation of outliergram see Arribas-Gil and Romo's article \cite{Arribas}. We can now indicate the functional observations (i.e., an index of a day in 2015), that are magnitude outliers, that is, days 164, 203, 222, 235, 242, 249, 255, 262, 263, 296, and 305 for service 1, they are mostly sunny days in the summer (9 out of 11) and 8 out of 11 days are Saturdays or Sundays; days 212 (Fri., July 31) and 306 (Mon., Nov. 2) for service 2 seem to show no pattern; as many as 94 days for service 3 are mostly Saturdays, Sundays and holidays, and surprisingly enough, the method does not indicate magnitude outliers for service 4. The method does not indicate functional magnitude outliers for the whole service as well.
\\ The functional shape outliers are days 266 (Wed., Sept. 23), 279 (Tue., Oct. 6), and 352 (Fri., Dec. 18) for service 1 and we conclude they show no common pattern; days 143 (Sat., May 23), 200 (Sun., July 19), and 358 (Christmas Eve - the most celebrated day in the whole year in Poland) for service 2; days 12, 101, 112, 120, 122, 136, 157, 164, 178, 181, 197, 198, 205, and 206 for service 3 (note that 8 out of 14 days are holidays); days 4, 8, 10, 46, 53, 54, 57, 58, 63, 66, 67, 68, 76, 81, 82, 86, 94, 95, 100, 117, 121, 136, 139, 151, 162, 186, 199, 200, 205, 209, 221, 226, 245, 250, 275, 294, 296, 297, 311, 313, 317, 341, 342, and 352 for service 4. Note that 18 out of 44 days are holidays in Poland and additionally 15 out of 44 days are days preceding or following holidays. Finally, days 10, 46, 66, 163, 205, and 336  (3 out of 6 days are weekends in winter and 3 out of 6 days are working days) are shape outliers for the whole service.
The performance of the aggregated median predictor, described in Section 3, applied to the analyzed number of Internet services users may be deduced from Figures \ref{fig:13}, \ref{fig:15} and \ref{fig:17}. The solid lines representing functional median in functional boxplots should be close to zero function (median-unbiasedness), whereas volumes of boxes should be relatively small (surrogate of effectiveness).
\\ Alternatively, the Shang and Hyndman's method can be used in order to obtain a forecast--a moving window of length equal to 10 is applied as well. 
We compared our method to the Shang and Hyndman's method \cite{Shang}. The performance of their predictor is presented on Figure \ref{fig:14}.\\ 
To sum up, we have compared quality of our forecast with the forecast of Shang and Hyndman \cite{Shang} through the comparison of sum of the differences between the observed curves and of the forecasted curves. We have empirically compared values of differences between the observed curves and the curves forecasted with both methods. We have also compared median absolute deviation (MAD) of the integrated differences between the observed curves and of the forecasted curves.
\\ In Figure \ref{fig:15}, we present four boxplots for the hourly average sum of the differences between the observed curves and the curves forecasted with the double functional median method and a functional boxplot for the forecasts of the number of service users.
\\ In Figure \ref{fig:16}, we present four boxplots for the median of sum of squares of the differences between the observed curves and the curves forecasted with the double functional median method.
\\ In Figure \ref{fig:17}, we present five functional boxplots for the values of differences between the observed curves and the curves forecasted for four subservices users forecasted  with the aggregated functional median method, whereas in Figure \ref{fig:18}, we present five functional boxplots for the  number of differences between the observed curves and the curves forecasted for four subservices users forecasted  with the Shang and Hyndman's method.


Comparing functional boxplots for the differences between the curves observed and curves predicted (see Figure \ref{fig:8}) has been conducted to compare "effectiveness" of our method with that of Shang and Hyndman's method.
Table \ref{tab:1} contains a comparison of our forecasts with that of Shang and Hyndman's-MAD has been calculated for the four subservices and for the whole service. We conclude that our methods works better for every single subservice, whereas Shang and Hyndman's method seems to be better in the whole service forecasting.
\begin{table}
\caption{Estimators quality comparison--MAD calculated for the Internet service users.}
\label{tab:1} 
\centering
\begin{tabular}{l|lllll}\hline
\noalign{\smallskip}
Predictor & sub-service1 & sub-service2& sub-service3& sub-service4& whole service\\\hline
S\&H & 75691911& 93756602& 53231453& 82312911 & 324824422\\
Our forecasts & 38363817& 42643286& 21726639& 37247856&1198788725\\
\hline
\end{tabular}
\end{table}

\section{Conclusions}
In this paper we have proposed a new forecasting method for the hierarchical functional time series. Our proposal comprises of an iterative calculation of moving functional medians induced by a functional depth. We begin from a bottom level of the hierarchy and we move upward the hierarchy computing relevant moving functional medians from the lower level medians until reaching the top level.\\
Taking into account the results of the simulation studies and empirical investigation, we conclude that our proposal is robust to both shape and magnitude outliers in cases of i.i.d. and weak temporal dependent models having strong merit justification. Our proposal may be easily adjusted to cope with specific functional outliers by applying different functional depths. We recommend using the $GBD$ for datasets consisting of the shape outliers and the $MBD$ for datasets consisting of the magnitude outliers.
\\ Taking into account the analysis of the empirical example, we conclude that our method works better for every single sublevel of hierarchy, whereas Shang and Hyndman's method seems to be better in the top level forecasting. Using our method, we were able to correctly predict the behavior of a majority of the web portal users. Historical data that are at a decision maker's disposal may contain functional outliers and may be time dependent in a certain degree.
Shang and Hyndman \cite{Shang} make forecasts basing on nonrobust generalized least squares method but the method seems to be robust globally, whereas our method is more robust locally, that is, on each sublevel. It should be stressed that our method is several times faster than Hyndman and Shang's method. We have faced a strong difficulties of the Hyndman and Shang's method in case of datasets of moderate sizes. This facts are very important in a context of applications of the proposal to e-economy analysis. 
\\ An implementation of the method may be easily prepared via simple R script appealing to \emph{DepthProc} R package. The package also consists of the considered empirical data set.
\\ \textbf{Acknowledgements}
\\ JPR and DM's research has been partially supported by the AGH local grant no. 11.11.420.004
and DK's research has been partially supported by the grant awarded to the Faculty of Management of CUE for preserving scientific resources for 2017 and 2018.

\end{document}